\documentstyle[12pt,a4]{article}
\def\L{ {\cal L}}

\def\mxth{\mathsurround=0pt }
\def\xversim#1#2{\lower2.pt\vbox{\baselineskip0pt \lineskip-.5pt
x  \ialign{$\mxth#1\hfil##\hfil$\crcr#2\crcr\sim\crcr}}}

\def\slash{\llap /}
\def\lagr{{\cal L}}
\newcommand{\thd}{{\theta_{\scriptscriptstyle\rm D}}}
\newcommand{\kild}[1]{{k^{#1}_{\scriptscriptstyle\rm D}}}

\renewcommand{\a}{\alpha}
\renewcommand{\b}{\beta}

\renewcommand{\d}{\delta}

\newcommand{\pa}{\partial}
\newcommand{\g}{\gamma}
\newcommand{\G}{\Gamma}

\newcommand{\e}{\epsilon}
\newcommand{\z}{\zeta}

\renewcommand{\l}{\lambda}
\renewcommand{\L}{\Lambda}
\newcommand{\m}{\mu}

\newcommand{\n}{\nu}

\newcommand{\s}{\sigma}

\renewcommand{\o}{\omega}

\newcommand{\Ka}{{K\"ahler}}
\renewcommand{\O}{{\Omega}}

\def\be{\begin{equation}}
\def\ee{\end{equation}}
\def\bea{\begin{eqnarray}}
\def\eea{\end{eqnarray}}

\newcommand{\ft}[2]{{\textstyle\frac{#1}{#2}}}

\newcommand{\eqn}[1]{(\ref{#1})}

\begin{document} 
\begin{titlepage}
\begin{center}
\hfill THU-99/23  \\
\hfill AEI-1999-18\\
\hfill SWAT-99/239 \\
\hfill YITP-99-51 \\[3mm]
\hfill {\tt hep-th/9909228}\\
\vskip 12mm

{\Large {\bf SUPERCONFORMAL HYPERMULTIPLETS}}

\vskip 10mm

{\bf{Bernard de Wit$^{a,b}$, Bas Kleijn$^a$ and 
Stefan Vandoren$^c$}}

\vskip 8mm

$^a${\em Institute for Theoretical Physics, 
Utrecht University,}\\
{\em 3508 TA Utrecht, Netherlands}\\
{\tt  B.deWit@phys.uu.nl, B.Kleijn@phys.uu.nl} \\[2mm]
$^b${\em Max Planck Institut f\"ur Gravitationsphysik, Albert Einstein
Institut,\\  
Am M\"uhlenberg 1, D-14476 Golm, Germany}\\[2mm]
$^c${\em C.N. Yang Institute for Theoretical Physics, SUNY, Stony
Brook.}\\
{\tt vandoren@insti.physics.sunysb.edu} 

\vskip 8mm

\end{center}

\vskip .2in

\begin{center} {\bf ABSTRACT } \end{center}
\begin{quotation}\noindent 
We present theories of $N=2$ hypermultiplets in four spacetime
dimensions that are
invariant under rigid or local superconformal symmetries. 
The target spaces of theories with rigid superconformal
invariance are  $(4n)$-dimensional {\it special} hyper-\Ka\
manifolds. Such manifolds can be described as cones  
over tri-Sasakian metrics and are locally the product of a flat
four-dimensional space and a quaternionic manifold. The latter
manifolds appear in the coupling of hypermultiplets to $N=2$
supergravity. We  
employ local sections of an Sp$(n)\times{\rm Sp}(1)$ bundle in 
the formulation of the Lagrangian and transformation rules, thus 
allowing for arbitrary coordinatizations of the hyper-K\"ahler
and quaternionic manifolds.
\end{quotation}

\vfill

September 1999\\

\end{titlepage}

\eject

\section{Introduction}
It is well known that in theories with rigid $N=2$  supersymmetry the 
hypermultiplet action takes the form of a supersymmetric sigma model
with scalars  that  parametrize a hyper-\Ka\ manifold \cite{AG-Fr}. In
the case of local supersymmetry the scalar 
fields parametrize a quaternionic manifold of negative curvature
\cite{BagWit}.  In this paper we study actions for hypermultiplets
invariant under rigid or local $N=2$ superconformal symmetries. This
study is both motivated by 
recent interest in superconformal theories \cite{sconf}  and by
efforts to find alternative and hopefully  more convenient
formulations of the hypermultiplet 
actions. 
The $N=2$ superconformal algebra in
four dimensions contains the bosonic subalgebra associated with
${\rm SO}(4,2)\times {\rm SU(2)} \times {\rm U(1)}$, together with 8
real supersymmetry and 8 real  `special' supersymmetry transformations,
called $Q$- and $S$-supersymmetry, respectively.
Requiring that the action is invariant under rigid superconformal
transformations leads to extra constraints on the target-space
geometry \cite{dubna}. For instance, these manifolds admit a so-called
hyper-K\"ahler potential whose derivative 
defines a conformal homothetic Killing vector and the three complex
structures rotate under the action of the SU(2) group, which must be
contained as a factor in the isometry group of the manifold. 
Spaces that satisfy these constraints will be called {\it
special} hyper-\Ka\ manifolds\footnote{
  Note that hyper-K\"ahler manifolds that are in the image of the
  {\bf c}-map are sometimes called special, because of the underlying
  special geometry features. We stress that the usage of the term
  {\it special} hyper-K\"ahler in this paper has no relationship to
  special geometry.}. 

By using the superconformal multiplet calculus \cite{DVV,DWLVP} one can then
obtain corresponding quaternionic sigma models coupled to $N=2$ 
supergravity. Because of the gauge degrees of freedom associated with
the dilatations and the SU(2) transformations, a $(4n)$-dimensional
special hyper-\Ka\ manifold leads to a $(4n-4)$-dimensional
quaternionic manifold.  At the time this
construction was applied to only flat 
hyper-\Ka\ spaces or hyper-\Ka\ quotients thereof. 
The coupling to supergravity then leads to a quaternionic projective
space and its quaternionic hyper-\Ka\ quotients \cite{DWLVP}. But
it has been known for some time that there exist 
quaternionic spaces that can couple to supergravity which are not in
this class but can be described in the context of the formalism of
\cite{BagWit}. Some of them have also been obtained explicitly
in the context of harmonic superspace \cite{BGIO}. Therefore it is
imperative to apply the superconformal approach to more
general special hyper-\Ka\ sigma models. This application is the main
topic of our paper, where, in order to avoid introducing an infinite
number of fields, we will no longer 
insist on off-shell supersymmetry for the hypermultiplets. 

Already quite some time ago the very same  option was discussed by Galicki
\cite{Galicki}. Rather than starting from the superconformally
invariant hypermultiplets, he described the geometry of these more
general hyper-\Ka\ spaces using a result of Swann
\cite{Swann}, who has proven that any quaternionic manifold has a
corresponding special hyper-\Ka\ manifold which admits a
quaternionically extended  
homothety and which has three complex structures that rotate under an 
isometric SU(2) action. And indeed, the hyper-\Ka\ manifolds
that he discusses have many properties in common with the
hypermultiplet actions discussed in \cite{dubna}. Moreover it is
known that a special hyper-\Ka\ manifold is a cone over a 
so-called tri-Sasakian manifold, so that there exists a beautiful relation
between quaternionic manifolds, tri-Sasakian manifolds and special
hyper-\Ka\ manifolds (for a recent review, see \cite{sasaki}). The
tri-Sasakian manifolds have also appeared recently in the context of
supergravity compactifications and the ADS/CFT correspondence
\cite{AFHS}.  

In this paper we follow the original superconformal approach and start
with the $(4n)$-dimensional  special hyper-\Ka\ manifolds as they were
formulated in \cite{dubna}. 
We establish that these spaces are indeed cones over
$(4n-1)$-dimensional tri-Sasakian spaces (this feature was also
discussed in \cite{GibbonsRych}). The special hyper-\Ka\ manifolds
have only a restricted holonomy group contained in Sp$(n-1)$; locally
they are a product of a flat four-dimensional space and a 
$(4n-4)$-dimensional quaternionic space. 
After gauging away the degrees of freedom associated with the
dilatations and the SU(2) transformations, the quaternionic space
remains when coupling to supergravity. We present the full Lagrangian
and transformation rules 
for the supersymmetric nonlinear sigma models based on special
hyper-\Ka\ spaces, including the option of gauged
isometries. Furthermore we construct local  ${\rm 
Sp}(n)\times {\rm Sp}(1)$ sections of  
the so-called associated quaternionic bundle which is
known to exist for any special hyper-\Ka\ manifold \cite{Swann}.
It turns out that the use of these sections greatly simplifies the
formulation of the transformation rules 
and the Lagrangian. In this way our general results remain closely
in line with the results of \cite{DWLVP}; the formulae are identical
up to modifications by connections and covariant tensors. When the
sections are trivial, so that the 
connections can be put to zero and the tensors become constant, they
can be identified with the hypermultiplet scalar fields and 
one directly recovers the results of \cite{DWLVP}.  
Guided by supersymmetry we thus make contact with the mathematical
results quoted above and we construct the general  action and
transformation rules in a new form. 

The last topic is to couple these supersymmetric nonlinear sigma
models to supergravity, using the conformal multiplet calculus. In
addition to presenting the corresponding field theory, we exhibit
how the quaternionic manifold emerges in the coupling. This
manifold can now be encoded in terms of ${\rm Sp}(n)\times
{\rm Sp}(1)$ sections that are projective with respect to
quaternionic multiplication.

Our results could facilitate the study of 
type-II string compactifications on Calabi-Yau three-folds.
These lead to four-dimensional models with both vector multiplets  and
hypermultiplets. 
While the moduli space of the vector multiplet scalars is described
in terms of a special \Ka\ geometry and is well understood, much less is 
known about the full quaternionic hypermultiplet moduli space.
It is known that at string tree level the quaternionic manifolds are 
obtained from a special \Ka\ manifold via the {\bf c}-map \cite{cfg}.
One would like to understand what the corrections are to the classical
hypermultiplet moduli space coming from both string perturbation theory
and non-perturbative effects \cite{quat-quantum}. With rigid
conformal symmetry, the results of this paper could also be helpful in the
description of cone branes \cite{AFHS}. 

This paper is organized as follows. In section~2 we briefly summarize
some essential features of hypermultiplet Lagrangians with gauged
target-space isometries. For hypermultiplets there exists no 
unconstrained off-shell formulation in terms of a finite number of
degrees of freedom, hence the supersymmetry algebra will only
be realized up to the field equations of the hypermultiplet
fermions. This is in contrast with the vector multiplets, introduced to
gauge the isometries,  and the superconformal theory itself, for which
off-shell formulations exist. As a result of the latter, the algebra of
gauged isometries and of the superconformal transformations, including
certain field-dependent structure constants, is
completely fixed and not affected by the presence of hypermultiplets.
Section~3 deals with
rigidly superconformal hypermultiplets, where we find 
the constraints on the hyper-\Ka\ manifold imposed by superconformal
invariance. The first subsection defines the superconformal
transformation rules, the second one deals with the hyper-\Ka\
potential and the construction of local  ${\rm 
Sp}(n)\times {\rm Sp}(1)$ sections, and the third one
gives the Lagragian and the transformation rules. 
The geometry of special hyper-\Ka\ manifolds is explained in
section 4. We first discuss the cone structure of these hyper-\Ka\
manifolds which lead to 
a tri-Sasakian space. The latter is indeed  an Sp(1) fibration over a
smaller space, which we prove to be quaternionic. This quaternionic
space couples to supergravity, as we then show in section~5. Here we
present the action for the hypermultiplets associated with a special
hyper-\Ka\ target space coupled to conformal supergravity and exhibit
how the target-space metric becomes quaternionic. 

\section{Preliminaries}
\setcounter{equation}{0}
In this section we summarize hypermultiplet Lagrangians in flat
spacetime. As is well known, these constitute $N=2$ supersymmetric
nonlinear sigma models with a hyper-\Ka\ target space
\cite{BagWit}. The holonomy group is contained in  Sp$(n)$ and it is this
group that is relevant for the hypermultiplet fermions. 
In the first subsection we discuss the supersymmetry transformations,
the Lagrangian and the target-space geometry. In a second subsection
we present possible extensions related to gauged 
target-space isometries, which will involve couplings to vector
multiplets associated with the gauge algebra. 

\subsection{Hypermultiplet nonlinear sigma models} 
We will base ourselves on the formulation of hypermultiplet
Lagrangians of \cite{DDKV}. With respect to the results of 
\cite{BagWit} this 
formulation differs in that it incorporates both a metric
$g_{AB}$ for the hyper-K\"ahler target space and a metric
$G_{\bar \a\b}$ for 
the fermions. Here we assume that the $n$ hypermultiplets are
described by $4n$ real scalars $\phi^A$, $2n$ positive-chirality  
spinors $\zeta^{\bar \a}$ and $2n$ negative-chirality spinors 
$\zeta^\a$. Hence target-space indices $A,B,\ldots$ take values
$1,2,\ldots, 4n$, and the indices $\a,\b, \ldots$ and $\bar\a,\bar\b,
\ldots$ run from 1 to $2n$. The chiral and antichiral spinors are
related by complex 
conjugation (so that we have $2n$ Majorana spinors) under which 
indices are converted according to $\a\leftrightarrow \bar \a$,
while SU(2) indices $i,j, \ldots=1,2$ are raised and lowered. An explicit
fermionic metric $G_{\bar\a\b}$ can be avoided as it can 
always be converted to a constant 
diagonal matrix by a similarity transformation. But retaining a
fermionic metric is, for example, important in obtaining transparant
transformation rules under symplectic transformations 
induced by the so-called $\bf c$-map from the electric-magnetic
duality transformations on a corresponding theory of vector
multiplets. In formulations based on $N=1$ superfields (such as
in \cite{HKLR}) one naturally has a fermionic metric but of a special
form. 

The Lagrangian and transformation rules are subject to a number of
equivalence transformations, two of which are associated with the target  
space. One set consists of the target-space diffeomorphisms 
$\phi\to \phi^{\prime}(\phi)$. The other refers to reparametrizations of 
the fermion `frame' of the form $\zeta^\a\to  S^\a{}_{\!\b} (\phi)\, \z^\b$, 
and corresponding redefinitions of other quantities carrying indices 
$\a$ or $\bar\a$. For example, the fermionic metric transforms as 
$G_{\bar\a\b} \to [\bar S^{-1}]^{\bar\g}{}_{\bar\a}\,
[S^{-1}]^\d{}_{\b}\, G_{\bar\g\d}$. 
There are connections, $\G_A{}^{\!\a}{}_{\!\b}$, associated with these
fermionic redefinitions, which appear in the Lagrangian and
supersymmetry transformation rules. Finally, there are chiral
SU$(2)\cong {\rm Sp}(1)$ redefinitions of the supercharges, which in the
rigidly supersymmetric case must be constant and are therefore
trivial. In the locally supersymmetric case this will be different and
in the latter part of this paper we will have to deal with local SU(2). 

The supersymmetry transformations are parametrized in terms of 
certain $\phi$-dependent quantities $\g^A$ and $V_A$ according to 
\bea
\d_{\scriptscriptstyle\rm Q}\phi^A&=& 2( \g^A_{i\bar\a} \,\bar\e^i 
\zeta^{\bar \a} +  
\bar\g^{Ai}_{\a} \,\bar\e_i \zeta^\a )\,,\nonumber\\
\d_{\scriptscriptstyle\rm Q}\zeta^\a  &=&  V_{A\,i}^\a \,\pa\slash\phi^A\e^i 
-\d_{\scriptscriptstyle\rm Q}\phi^A\, \G_{A}{}^{\!\a}{}_{\!\b}\,
\zeta^\b \,, \nonumber\\
\d_{\scriptscriptstyle\rm Q}\zeta^{\bar \a}&=& \bar V_A^{i\bar\a} 
\,\pa\slash\phi^A\e_i -\d_{\scriptscriptstyle\rm Q}\phi^A\,
\bar\G_{A}{}^{\!\bar\a}{}_{\!\bar \b} \,\zeta^{\bar \b}  \,.  
\label{4dhsusy} 
\eea
In principle $\g^A$ and $V_A$ each denote $(4n)\times(4n)$ complex
quantities, but as we shall see below, these quantites are related and
satisfy a pseudoreality condition. As it turns out they will play the
role of the quaternionic (inverse) vielbeine of the target space. 
Observe that the supersymmetry variations are consistent with a U(1) chiral
invariance under which the scalars remain invariant, while the fermion
fields and the 
supersymmetry transformation parameters transform. This group will be
denoted by ${\rm U(1)}_{\rm R}$ to indicate that it is a subgroup  
of the automorphism group of the supersymmetry algebra. In
section~3 we will see that this U(1) will correspond to one of the
conformal gauge transformations. However, 
for generic $\g^A$ and $V_A$, the SU(2)$_{\rm R}\cong {\rm Sp}(1)$
part of the automorphism group cannot be realized consistently on the
fields. This would require the presence of an SU(2) isometry in the
target space. 
In the above, we merely used that $\z^\a$ and $\z^{\bar \a}$ are related 
by complex conjugation. 

The Lagrangian takes the following form
\be
 \lagr= -\ft12 g_{AB}\,\pa_\m\phi^A\pa^\m\phi^B  
-G_{\bar \a \b}( \bar\zeta^{\bar \a} D\!\slash \,\zeta^\b +  
\bar\zeta^\b D\!\slash \,\zeta^{\bar\a}) -\ft14  W_{\bar
\a\b\bar\g\d}\, \bar \zeta^{\bar \a}  
\g_\m\zeta^{\b}\,\bar \zeta^{\bar \g} \g^\m\zeta^\d 
\,, \label{4dhlagr1}
\ee
where we employed the covariant derivatives
\be
D_\m \zeta^\a= \pa_\m \zeta^\a + \pa_\m\phi^A\, 
\G_{A}{}^{\!\a}{}_{\!\b} \,\zeta^\b\,, 
\quad
D_\m \zeta^{\bar\a}= \pa_\m \zeta^{\bar \a}  +\pa_\m\phi^A\,\bar 
\G_{A}{}^{\!\bar\a}{}_{\bar \b} \,\zeta^{\bar \b} \,.
\ee
Besides the Riemann curvature $R_{ABCD}$ we will be dealing with
another curvature $R_{AB}{}^{\!\a}{}_{ \b}$ associated with the
connections $\G_{A}{}^{\!\a}{}_{\!\b}$, which takes its values in
$sp(n)\cong usp(2n;{\bf C})$. The tensor $W$ is defined by  
\be
W_{\bar \a \b \bar \g\d} = 
R_{AB}{}^{\!\bar\e}{}_{\bar \g} \,\g^A_{i\bar\a}\,\bar \g^{iB}_\b\, 
 G_{\bar \e\d}  = \ft12 R_{ABCD} \,\g^A_{i\bar\a}\,\bar \g^{iB}_\b\, 
\g^C_{j\bar\g} \,\bar \g^{jD}_\d\, , \label{defW}
\ee
and will be discussed shortly in more detail.

The target-space metric $g_{AB}$, the tensors
$\g^A$, $V_A$ and the fermionic hermitean metric $G_{\bar\a\b}$ (i.e.,
satisfying $(G_{\bar \a\b})^\ast = G_{\bar \b\a}$) are all
covariantly constant with respect to the Christoffel connection
and the connections $\G_{A}{}^{\!\a}{}_{\!\b}$. Furthermore
we note the following relations,
\bea
&&\g^A_{i\bar\a} \,\bar V_B^{j\bar\a} + \bar \g^{A\,j}_\a \,V_{B\,
i}^\a  = \d^j_i\,\d^A_B\,, \label{clifford}  \nonumber \\
&&g_{AB}\, \g^B_{i\bar \a} = G_{\bar\a\b}\, V_{A\,i}^\b 
\,,   \qquad
\bar V^{i\bar \a}_A \, \g^A_{j\bar \b} = \d^i_j\, 
\d^{\bar \a}_{\,\bar \b}\,. \label{inverse}
\eea
{}From them one derives a number of useful relations, such as 
\be
  \bar \g_{A\a}^j \,V_{Bi}^\a = \gamma_{Bi\bar\alpha}
  \,\bar V^{j\bar \a}_A = - \bar \gamma^j_{B\alpha}
  \,V^\a_{i A} + \d^j_i \,g_{AB}\,.
\ee
The following three bilinears define antisymmetric covariantly constant 
target-space tensors, 
\be
J^{ij}_{AB} = \g_{Ak\bar\a}\,\varepsilon^{k(i}\bar V^{j)\bar\a}_B\,,
\ee
that span the complex structures of the
hyper-K\"ahler target space. They satisfy 
\be
(J_{ij})_{AB} \equiv (J^{ij}_{AB})^\ast =
\varepsilon_{ik}\varepsilon_{jl}\,J^{kl}_{AB}\,,\qquad 
J^{ij C}_{\,A} \,J^{kl}_{CB} = \ft12
\varepsilon^{i(k}\varepsilon^{l)j}\, g_{AB}
+ \varepsilon_{~}^{(i(k} \, J^{l)j)}_{AB}\,.\label{JJ}
\ee
In addition we note the following useful identities, 
\be
\g_{Ai\bar\a}\, \bar V^{j\bar \a}_B= \varepsilon_{ik} J^{kj}_{AB} +
\ft12 g_{AB}\, \d^j_i\,,\qquad J_{AB} ^{ij}\, \gamma_{\bar \a
k}^B= -\d^{(i}_k \varepsilon_{~}^{j)l}\,\gamma_{Al\bar\a}\,. 
\label{bilinear-id}
\ee
We also note the existence of covariantly constant 
antisymmetric tensors,  
\be
\O_{\bar\a\bar\b} =\ft12  \varepsilon^{ij}\,g_{AB}\, \g^A_{i\bar 
\a}\,\g^B_{j\bar \b}\,,
\quad 
\bar\O^{\bar\a\bar \b} =\ft12  \varepsilon_{ij}\,g^{AB}\, \bar 
V_A^{i\bar \a}\,\bar V_B^{j\bar \b}\,,
\ee
satisfying $\O_{\bar\a\bar\g}\,\bar \O^{\bar\g\bar\b} = -\d_{\bar 
\a}{}^{\!\bar \b}$. Their complex conjugates satisfy
\begin{equation}
\bar \O_{\a\b}\equiv (\O_{\bar \a \bar \b})^*=G_{\bar \g \a}{\bar \O}^{\bar
\g \bar \d}G_{\bar \d \b}\ .
\end{equation}
The tensor $\Omega$ can be used to define a reality condition on $V$
and $\gamma$,
\be
\varepsilon_{ij}\,\Omega_{\bar\a\bar\b}\,\bar V^{j\bar\b}_A = g_{AB}
\,\gamma^B_{i\bar\a} = G_{\bar\a\b}\,V^\b_{A\,i}\,. \label{pseudoreal} 
\ee
This equation leads to 
\be
g^{AB}\, V_{Ai}^\a\,V_{Bj}^\b = \varepsilon_{ij} \,
\Omega^{\a\b}\,,\qquad  g_{AB}\, \g^A_{i\bar\a}\,\g^B_{j\bar \b} =
\varepsilon_{ij} \, \Omega_{\bar\a\bar \b}\,  .
\ee
Another convenient identity is given by 
\be
\bar V_A^{i\bar \a}\,\Omega_{\bar\a\bar\b}\, \bar V_B^{j\bar \b} = \ft12
\varepsilon^{ij} \,g_{AB} - J_{AB}^{ij} \;. 
\ee

The existence of the covariantly constant tensors implies a variety 
of integrability conditions which have a
number of consequences for the various curvature tensors
\cite{BagWit,DDKV}. First of all the covariant constancy of
$\bar\gamma^A$ implies
\be
R_{ABCD} \,\bar\g^{Ci}_\a\,\bar\g^{Dj}_\b = - 
\varepsilon^{ij} \,\bar\Omega_{\a\g}\,R_{AB}{}^{\!\g}{}_{\!\b}
\,. \label{integrab1} 
\ee
Observe that the right-hand side is manifestly antisymmetric in $[ij]$
and symmetric in $(\a\b)$. This implies that the Riemann tensor can be
written with tangent-space indices according to 
\be
R_{ABCD} \,\bar\g^{Ai}_\a\,\bar\g^{Bj}_\b
\,\bar\g^{Ck}_\g\,\bar\g^{Dl}_\d  = \ft12 \varepsilon^{ij} \,
\varepsilon^{kl} \,W_{\a\b\g\d}\,, \label{defW-symm}
\ee
where, as a result of the cyclicity property of the Riemann tensor,
$W_{\a\b\g\d}$ is symmetric in all four indices. This tensor is
linearly related to the tensor \eqn{defW} upon multiplication with the
tensors $G$ and $\Omega$. In terms of $W_{\a\b\g\d}$ the curvatures read
\bea
R_{ABCD} &=& \ft12 \varepsilon^{ij}\,\varepsilon^{kl}\,
V_{Ai}^\a\,V_{Bj}^\b\,V_{Ck}^\g\,V_{Dl}^\d \,W_{\a\b\g\d} \,,\nonumber\\
\bar\Omega_{\a\e}\,R_{AB}{}^{\!\e}{}_{\!\b} &=&-  \ft12 \varepsilon^{ij}
\,V_{Ai}^\g\,V_{Bj}^\d\,W_{\a\b\g\d}\,.
\eea

The above results are all derived from the requirement of
supersymmetry. To characterize the geometry of the target space, one
could start from the nonsingular $\bar V_A^{i\bar\a}$ and a 
nonsingular skew-symmetric tensor $\Omega_{\bar\a\bar\b}$ that is
covariantly constant with respect to a symplectic connection 
$\bar\G_A{}^{\!\bar\a}{}_{\!\bar \b}$. Subsequently one notes that  
$\varepsilon_{ij}\,\Omega_{\bar\a\bar\b}\,\bar V^{j\bar\b}_A$ and the
inverse of $\bar V_A$, denoted by $\gamma^B_{i\bar\a}$, are linearly
related by a symmetric matrix $g_{AB}$. Requiring that this matrix is
real we can identify it with the target-space metric while the ensuing
reality constraint on the $V_A$ enables their identification  as the
corresponding quaternionic vielbeine. This information is sufficient for
deriving all the algebraic identities listed above. The vielbeine and
the symplectic connection then allow the 
definition of an affine target-space connection, with respect to which
the vielbeine are covariantly constant thus leading to a generalized
vielbein postulate. All of above results then follow upon
assuming that the target space has no torsion so that the affine
connection and the Christoffel connection coincide. 
\subsection{Gauged target-space isometries} 
The equivalence transformations of the fermions and the 
target-space diffeomorphisms do not constitute invariances of the 
theory. This is only the case when the metric $g_{AB}$ and 
the Sp($n$)$\times$Sp(1) one-form $V^\a_i$ (and thus the
related geometric quantities) are left invariant under 
(a subset of) them. Therefore these are related to isometries of the
hyper-K\"ahler space. We can then elevate such invariances to a group of
local (i.e. spacetime-dependent) transformations, by introducing the
required gauge fields in the form of vector multiplets. Such gauged
isometries have been studied earlier in the literature 
\cite{ST,HKLR,BGIO,DFF,ABCDFFM} and the purpose of our discussion here 
is to incorporate them into the formulation used in this paper.

We consider scalar fields transforming under a certain  
isometry (sub)group G characterized by a number of Killing vectors 
$k^A_I(\phi)$, with parameters $\theta^I$. 
Hence under infinitesimal transformations, 
\begin{equation}
\delta_{\rm G} \phi^A=g\,\theta^Ik_I^A(\phi)\ ,
\end{equation}
where $g$ is the coupling constant and the $k^A_I(\phi)$ satisfy 
the Killing equation, 
\begin{equation}
D_Ak_{IB} + D_B k_{IA}=0\,. \label{killing-eq}
\end{equation} 
The isometries constitute an algebra with structure
constants $f_{IJ}{}^{\!K}$,
\begin{equation}
k^B_I\partial_Bk^A_J-k^B_J\partial_Bk^A_I =-f_{IJ}{}^K\, k^A_K\ .
\label{killingclosure}
\end{equation} 
Our definitions are such that the gauge fields that are
needed 
once the $\theta^I$ become spacetime dependent, transform according to  
$\delta_{\rm G}W^I_\m=\partial_\mu\theta^I- gf_{JK}{}^I\,W^J_\mu\,
\theta^K$.
The Killing equation generally implies the following property 
\begin{equation}
D_{A}D_{B} k_{IC}  =  R_{BCAE}\,  k^{\,E}_I  \ . 
\label{symmetric}
\end{equation}

Quantities that carry Sp($n$) indices, such as $V^\a_{Ai}$, are
only required to be 
invariant under isometries up to fermionic equivalence 
transformations. Thus $-g(
k^B_I \,\pa_B V^\a_{Ai} + \partial_A k_I^B\, V^\a_{Bi})$ must be
cancelled by a suitable   
infinitesimal rotation on the index $\a$. 
Here we assume that the effect of  
the diffeomorphism is entirely compensated by a rotation that
affects the indices $\a$. In principle, one 
can also allow for a compensating Sp(1) transformation acting on the indices
$i,j,\ldots$. However, the latter transformations must be constant, so
they will generically not appear here. This is equivalent to requiring
that the isometry group will commute with supersymmetry. 

Let us parametrize the compensating transformation
acting on the Sp($n$) indices by 
$\delta_{\rm G} \z^\a= g[{t_I} - k_I^A\, \Gamma_A]^{\a}{}_{\!\b} \,\z^\b$, 
where the ($\phi$-dependent) matrices $t_I(\phi)$ remain to be
determined, 
\be
-k^B_I \,\pa_B V_{A\,i}^\a - \pa_A k^B_I \,V_{B\,i}^\a + 
(t_I -k^B_I\,\G_B)^{\a}{}_\b \, V_{A\,i}^{\!\b} =0\,. 
  \label{invariant-bein}
\ee
Obviously similar equations apply to the other geometric
quantities, but as those are not independent we do not need to
consider them. 
Using the covariant constancy of 
$V_A$, we derive from \eqn{invariant-bein},
\be
(t_I)^{\a}{}_{\!\b}\, V_{A\,i}^\b  = D_Ak_I^{\,B}\,
 V_{B\,i}^{\a} \,, \label{t-V-relation} 
\ee
so that 
\be
(t_I)^{\a}{}_{\!\b} = \ft12 V_{Ai}^{\a} \,
\bar\gamma^{Bi}_{\b}\, D_Bk_I^{\,A}\,. 
\ee
Target-space scalars will satisfy algebraic identities, such as 
\be
(\bar t_I)^{\bar\g}{}_{\!\bar\a} \, G_{\bar\g\b} 
+(t_I)^{\g}{}_{\!\b} \, 
G_{\bar\a\g}= (t_I)^{\bar\g}{}_{\![\bar\a} \, \Omega_{\bar\b]\bar\g} =
0\,. 
\ee
This establishes that the field-dependent matrices $t_I$ take 
values in $sp(n)$. From \eqn{killing-eq} and \eqn{symmetric}, it
easily follows that 
\begin{equation}
D_A t_I{}^{\!\a}{}_{\!\b}  =  k^{\,B}_I\, 
R_{AB}{}^{\!\a}{}_{\!\b} \,, \label{t-der}
\end{equation}
for any infinitesimal isometry. From the group property of the
isometries it follows that 
the matrices $t_I$ satisfy the commutation relation 
\be
[\,t_I ,\,t_J\,]^\a{}_{\!\b}   = f_{IJ}{}^K\, 
(t_K)^\a{}_{\!\b}  + k^A_I\,k^B_J\, 
R_{AB}{}^{\!\a}{}_{\!\b} \,,  \label{t-comm}
\ee
which takes values in $sp(n)$.
The apparent lack of closure represented by the presence of the
curvature term is related to the fact that the
coordinates $\phi^A$ on which the matrices depend, transform
under the action of the group. One can show that this result is
consistent with the Jacobi identity. 

Furthermore we derive from 
\eqn{invariant-bein} that the complex structures $J_{AB}^{ij}$ are 
invariant under the isometries, 
\be
k^C_I \pa_C J_{AB}^{ij} - 2 \pa_{[A}k^C_I \,J_{B]C}^{ij}  =
0\,. \label{triholo} 
\ee
This means that the isometries are {\it tri-holomorphic}. From
\eqn{triholo} one shows that $\pa_A(J^{ij}_{BC}\, k^C_I )
-\pa_B(J^{ij}_{AC}\, k^C_I )=0$, 
so that, locally, one can associate three Killing potentials (or
moment maps) $P^{ij}_I$ to every Killing vector, according to
\be
\pa_A P_I^{ij}  = J_{AB}^{ij} \,k^{\,B}_I \,. \label{moment}
\ee
Observe that this condition determines the moment maps up to a 
constant. Up to constants one can also derive the 
equivariance condition,
\begin{equation}
J^{ij}_{AB}\,k^A_Ik^B_J=-f_{IJ}{}^{\!K}\,P^{ij}_K\ , \label{equivariance}
\end{equation}
which implies that the moment maps transform covariantly under 
the isometries,
\be
\d_{\rm G}P_I^{ij} = \theta^J\,k_J^A\,\pa_A P^{ij}_I  = - 
f_{JI}{}^{\!K} \,P^{ij}_K\,\theta^J \, . \label{moment-isometry}
\ee

Summarizing, the invariance group of the isometries acts as follows,
\begin{equation}
\d_{\rm G}\phi = g\,\theta^I\,k_I^{\,A}\,,\qquad \delta_{\rm G} 
\z^\a=g\, 
(\theta^I{t_I})^{\a}{}_{\!\b}\,\z^\b - \d_{\rm G}\phi^A  
\Gamma_A{}^{\!\a}{}_{\!\b} \,\z^\b \,.\label{fermgaugetr}
\end{equation}
When the parameters of these isometries become spacetime 
dependent we introduce corresponding gauge fields and fully 
covariant derivatives, 
\bea
{\cal D}_\mu \phi^A = \partial_\m \phi^A - g W^I_\m \,k_I^A \,
, \quad
{\cal D}_\mu\z^\a =\partial_\mu \z^\a+\pa_\mu\phi^A\,
{\G_A}^{\!\a}{}_{\!\b}  
\z^\b -gW_\m{}^{\!\!\a}{}_{\!\!\b}\z^\b\, , \;
\eea
where $W_\m{}^{\!\a} {}_{\!\b} =W^I_\m\,(t_I)^{\a}{}_{\!\b}$. The 
covariance of ${\cal D}_\m\zeta^\a$ depends crucially on \eqn{t-der} and 
\eqn{t-comm}; after some calculation one finds 
\be
\delta_{\rm G} {\cal D}_\m \z^\a=g\, 
(\theta^I{t_I})^{\a}{}_{\!\b}\,{\cal D}_\m \z^\b - \d_{\rm G}\phi^A  
\Gamma_A{}^{\!\a}{}_{\!\b} \,{\cal D}_\m\z^\b \,. 
\ee
The gauge
fields $W^I_\mu$ are accompanied by complex scalars $X^I$, spinors
$\Omega_i^I$ and auxiliary fields $Y_{ij}^I$, constituting off-shell
$N=2$ vector multiplets. For our notation of vector multiplets,
the reader may consult \cite{DDKV}. 

The minimal coupling to the gauge fields requires extra terms in
the supersymmetry transformation rules for the hypermultiplet
spinors as well as in the Lagrangian, in order to regain $N=2$
supersymmetry. The extra terms in the transformation rules are
\be
\delta^\prime _{\scriptscriptstyle\rm Q}\z^\a = 
2gX^Ik_I^AV^\a_{Ai}\,\varepsilon^{ij}\e_j\,,\qquad  
\delta^\prime_{\scriptscriptstyle\rm Q}\z^{\bar \a}=
 2g{\bar X}^Ik_I^A{\bar V}^{{\bar \a} 
i}_{A}\,\varepsilon_{ij}\e^j\ . \label{susyferm} 
\ee
These terms can be conveniently derived by imposing the commutator of
two supersymmetry transformations on the scalars, as this
commutator should yield the correct field-dependent gauge
transformation.  

We distinguish three additional couplings to the Lagrangian. The
first one is quadratic in the hypermultiplet spinors and reads
\begin{eqnarray}
\lagr_g^{(1)} =g{\bar X^I}{\bar \g}^{Ai}_\a\e_{ij}{\bar
\g}^{Bj}_\b\,D_Bk_{AI}\,{\bar \z}^\a\z^\b + \mbox{h.c.}
=2g{\bar X}^I{t_I}^{\!\gamma}{}_{\!\a}\,\bar \O_{\b\gamma}\,{\bar 
\z}^\a\z^\b+ \mbox{h.c.} \ .\label{S1}
\end{eqnarray}
The second one is proportional to the vector multiplet spinor
$\Omega^I$ and takes the form
\begin{equation}
\lagr_g^{(2)} =-2gk^A_IV_{Ai}^\a \bar\Omega_{\a\b}\,{\bar 
\z}^\b\O^{Ii} + 
\mbox{h.c.} =2gk^A_I{\bar \g}^i_{A\a}\e_{ij}\,{\bar
\z}^\a \Omega^{Ij}  + \mbox{h.c.} \ .\label{S2}
\end{equation}
Finally there is a potential given by
\begin{equation}
\lagr_{g}^{\rm scalar} =-2g^2k^A_Ik^B_J\,g_{AB}\,X^I{\bar X^J} + g \,
P^{ij}_I\, Y^I_{ij}\ ,\label{pot}
\end{equation}
where $P^{ij}_I$ is the triplet of moment maps on the
hyper-K\"ahler space. These terms were determined both from imposing
the supersymmetry algebra and from the invariance of the action. 
To prove \eqn{pot}, one has to make use of the equivariance
condition \eqn{equivariance}. Actually, gauge invariance, which 
is prerequisite to supersymmetry, already depends on 
\eqn{moment-isometry}.

\section{Rigidly superconformal hypermultiplets}
\setcounter{equation}{0}
In this section we determine the restrictions
on the hyper-K\"ahler geometry that follow from imposing invariance
under rigid superconformal transformations.  As we already mentioned
in section~1, the corresponding spaces,  called {\it special} 
hyper-\Ka\ manifolds, have an intriguing geometrical structure. In
section~5  we will obtain 
the coupling of hypermultiplets to conformal supergravity. A crucial
element in the construction of this coupling is that the full
superconformal theory is known in an off-shell form, so that the
superconformal algebra remains unaffected in the 
presence of matter fields. Our goal is more modest in this section
where we only consider rigid superconformal transformations. This
aspect does not play a role for the derivation of the superconformal 
transformations on the hypermultiplets and the results of this section
describe the situation that would arise when freezing all the fields
of conformal supergravity to zero in a flat spacetime metric. In that
case the superconformal transformations acquire an explicit but fixed
dependence on the spacetime coordinates parametrized by a finite number of
spacetime-independent parameters (this is explained, for instance, in
\cite{DW}). 

In the first subsection we impose the superconformal
algebra on the fields and find the transformation rules as well as a
number of important results for the complex structures and the moment
maps associated with possible isometries. In the second subsection we
derive the existence of a hyper-K\"ahler potential and reformulate the
theory in terms of local sections of an Sp$(n)\times {\rm Sp(1)}$
bundle. Then, in the third subsection, we present the Lagrangian and
the transformation rules in terms of these local sections.
\subsection{Superconformal transformations}
We start by implementing the $N=2$ superconformal algebra \cite{DVV} 
on the hypermultiplet fields. We assume that the scalars are invariant 
under special conformal and special supersymmetry 
transformations, but they transform under $Q$-supersymmetry and 
under the additional   
bosonic symmetries of the superconformal algebra, namely chiral 
[SU(2)$\times$U(1)]$_{\rm R}$ and dilatations denoted by $D$. At 
this point we do not assume  
that these transformations are symmetries of the action and we 
simply parametrize them as follows,
\be
\d\phi^A = \thd\, \kild{A}(\phi) + \theta_{\scriptscriptstyle\rm U(1)}\, 
k^A_{\scriptscriptstyle\rm U(1)}(\phi)  
+(\theta_{\scriptscriptstyle\rm SU(2)})^i{}_k \,\varepsilon^{jk} 
k^A_{ij}(\phi) \,,
\ee
where the $k^A$ are left arbitrary. Note that  $k^A_{ij}(\phi)$ 
is assigned to the same symmetric pseudoreal representation of 
SU(2) as the complex structures, while 
$\theta_{\scriptscriptstyle\rm SU(2)}$ is antihermitean and 
traceless.  

An important difference with the situation described in the 
previous section,  is that in the conformal superalgebra 
the dilatations and chiral transformations do not appear in the 
commutator of two $Q$-supersymmetries, but in the commutator of a 
$Q$- and an $S$-supersymmetry. To evaluate the $S$-supersymmetry 
variation of the fermions, we assume that $\d_{\scriptscriptstyle\rm 
S}\phi^A = \d_{\scriptscriptstyle\rm K} \zeta^\a=0$   
and covariantize the derivative in the fermionic transformations 
with respect to dilatations. Subsequently we impose the 
commutator, $[\,\delta_{\scriptscriptstyle\rm 
K}(\Lambda_{\scriptscriptstyle\rm K}),
\delta_{\scriptscriptstyle\rm Q}(\epsilon)\,] =  
-\delta_{\scriptscriptstyle\rm S}(\,\rlap/\!\L_{\scriptscriptstyle\rm
K} \epsilon)$ 
on the spinors. This expresses the $S$-supersymmetry variations 
in terms of $\kild{A}$,
\be 
  \delta_{\scriptscriptstyle\rm S}(\eta)\,\z^{\a} = {V}^{\a}_{i\,A} \, 
\kild{A}\,\eta^i\,,\qquad 
  \delta_{\scriptscriptstyle\rm S}(\eta)\,\z^{\bar\a} = 
{\bar{V}}^{i\,\bar\a}_A \,  \kild{A}\,\eta_i.
  \label{eq:dSTwo}
\ee
With this result we first evaluate the commutator of an $S$- and a 
$Q$-supersymmetry transformation on the scalars. This yields
\be 
{}[\,\d_{\scriptscriptstyle\rm S}(\eta), 
\d_{\scriptscriptstyle\rm Q}(\e)\,] \,\phi^A = (\bar \e^i\eta_i   
+\bar\e_i\eta^i) \, \kild{A} + 2J_{ik}{}^{\!\!A}{}_{\!B} \,
\varepsilon^{kj}\,
(\bar \e^i\eta_j -\bar\e_j\eta^i) \,\kild{B}\,. 
\ee
This result can be confronted with the corresponding expression from  the 
$N=2$ superconformal algebra, which reads 
\bea
{}  [\,\delta_{\scriptscriptstyle\rm S}(\eta), 
\delta_{\scriptscriptstyle\rm Q}(\epsilon)\,] &=&  
    \delta_{\scriptscriptstyle\rm M} 
(2\bar{\eta}^i\s^{ab}\epsilon_i + {\hbox{h.c.}}) 
    +\delta_{\scriptscriptstyle\rm D} (\bar{\eta}_i \epsilon^i + 
{\hbox{h.c.}})\nonumber\\ 
  & &+\delta_{\scriptscriptstyle\rm U(1)} (i\bar{\eta}_i 
\epsilon^i +{\hbox{h.c.}}) 
    +\delta_{\scriptscriptstyle\rm SU(2)} (-2  \bar{\eta}^i \epsilon_j 
-{\hbox{h.c.} \, ; \, {\rm traceless})}\,.   
  \label{eq:SQComm}
\eea
Comparison thus shows that $k^A_{\scriptscriptstyle\rm U(1)}$ 
vanishes and that the SU(2) vectors satisfy
\be
k^A_{ij} =  J_{ij}{}^{\!A}{}_{\!B}\, \kild{B}\,.  
\label{su(2)-vector}
\ee
Now we proceed to impose the same commutator on the fermions, 
where on the right-hand side we find a Lorentz transformation, a 
U(1) transformation and a dilatation, if and only if we assume the 
following condition on $\kild{A}$,
\be
D_A\kild{B} = \d^B_A\,.  \label{homo-vector}
\ee
The geometric significance of these results will be discussed in later
subsections. Here we note that \eqn{homo-vector} suffices to show
that the kinetic term of the scalar fields is invariant under
dilatations, provided one includes a spacetime metric or, in  
flat spacetime, includes corresponding scale transformations of 
the spacetime coordinates. Nevertheless, observe that $\kild{A}$ 
is {\it not} a Killing vector of the hyper-K\"ahler space, although it
still satisfies \eqn{symmetric}, but an example of a conformal
homothetic Killing vector. Another consequence is that  
the SU(2) vectors $k^A_{ij}$, as expressed by \eqn{su(2)-vector}, 
are themselves Killing vectors,  
because their derivative is proportional to the corresponding 
antisymmetric  complex structure 
\be
D_Ak_B^{ij} = - J_{AB}^{ij}\,. \label{D-su(2)}
\ee
{}From this it follows that the K\"ahler two-forms are exact, provided
that the Killing vectors are globally defined. The
product rule of the SU(2) 
Killing vectors can now be worked out and one finds 
\be
k^{B\,ij}\,\pa_B k^{A\,kl} -  k^{B\,kl}\,\pa_B k^{A\,ij} =
2\, k^{A\,(i(k}\,\varepsilon^{l)j)}  \,,  
\ee
which is indeed in accord with the SU(2) structure constants. 

{}From the $[\d_{\scriptscriptstyle\rm S},\d_{\scriptscriptstyle\rm 
Q}]$ commutator we also establish the fermionic 
transformation rules under the chiral transformations and the 
dilatations,
\bea 
  \delta_{\scriptscriptstyle\rm SU(2)}\,\zeta^{\a} + 
\delta_{\scriptscriptstyle\rm SU(2)}\phi^A\,
    \Gamma_A{}^{\!\a}{}_{\!\b}\,\zeta^\b &=& 0\, ,\nonumber \\
 \delta_{\scriptscriptstyle\rm U(1)}\zeta^\a + 
\delta_{\scriptscriptstyle\rm U(1)}\phi^A \, 
\Gamma_A{}^{\!\a}{}_{\!\b}\,\zeta^\b
&=& -\ft12i\,\theta_{\scriptscriptstyle\rm U(1)}\zeta^\a\,,  
\nonumber \\
  \delta_{\scriptscriptstyle\rm D}\zeta^\a + 
\delta_{\scriptscriptstyle\rm D}\phi^A\,
    \Gamma_A{}^{\!\a}{}_{\!\b}\zeta^\b &=& \ft32\,
\theta_{\scriptscriptstyle\rm D}\,\zeta^\a\,.
\eea
Note that the U(1) transformation further simplifies because 
$\delta_{\scriptscriptstyle\rm U(1)}\phi^A=0$. 

To establish that the model as a whole is now invariant under the 
superconformal transformations it remains to be shown that the 
tensor $V^\a_{Ai}$ is invariant under the diffeomorphisms 
generated by $k_{ij}^A$, $k_{\scriptscriptstyle\rm U(1)}^A$ and 
$k_{\scriptscriptstyle\rm D}^A$ up to compensating transformations that act on 
the Sp($n$)$\times$Sp(1) indices in accordance with the 
transformations of the $\zeta^\a$ given above and the symmetry 
assignments of the supersymmetry parameters $\e^i$. To emphasize 
the systematics we ignore the fact that $k_{\scriptscriptstyle\rm 
U(1)}^A$ actually vanishes and we write 
\bea
- k_{kl}^B\, \pa_B V^\a_{Ai} - \pa_Ak^B_{kl}\,  V^\a_{Bi} -  
k^B_{kl}\,\Gamma_B{}^{\!\a}{}_{\!\b} V^\b_{Ai} + 
[- \d_{(k}^j\varepsilon^{~}_{l)i}]\, 
V^\a_{Aj} &=& 0\,,
\nonumber\\
- k^B_{\scriptscriptstyle\rm U(1)} \pa_B V^\a_{Ai} - 
\pa_Ak^B_{\scriptscriptstyle\rm U(1)} \, V^\a_{Bi} +
[- \ft12 i\,\d^\a_\b- k^B_{\scriptscriptstyle\rm U(1)} \,
\Gamma_B{}^{\!\a}{}_{\!\b}]  V^\b_{Ai} +  [\ft12 i\,\d^j_i]\, 
V^\a_{Aj} &=& 0\,, 
\nonumber\\
- k^B_{\scriptscriptstyle\rm D} \,\pa_B V^\a_{Ai} - 
\pa_Ak^B_{\scriptscriptstyle\rm D} \,V^\a_{Bi} +
[\ft32 \d^\a_\b-  \kild{B}\,\Gamma_B{}^{\!\a}{}_{\!\b} ]\,  
V^\b_{Ai} + [- \ft 12\, \d^j_i]\, V^\a_{Aj} &=& 0\,. \;\;{~}
\label{comp-sc}
\eea
In these equations the first two terms 
on the left-hand side represent the effect of the isometry or dilatation, the 
third term represents a uniform scale and chiral U(1) 
transformation on the indices associated with the Sp($n$) tangent 
space, and the last terms represent an SU(2), a U(1) and a scale 
transformation, respectively, on the indices associated with 
Sp(1). Eq. \eqn{comp-sc} should be regarded as a 
direct extension of \eqn{invariant-bein}. 

We close with a few comments. First of all, the SU(2) isometries 
induce a rotation on the complex structures, 
\be
k_{kl}^C\,\pa^{~}_C J^{ij}_{AB} - 2\pa_{[A}k^C_{kl}\,
J_{B]C}^{ij} = -2 J_{klC[A} \,J^{ijC}_{\,B]}= 2 \d_{(k}^{(i} \, 
\varepsilon^{~}_{l)m}\,J ^{j)m}_{AB}\,,
\ee
as should be expected. 
Under dilatations, the \Ka\ two-forms $J_{AB}$ scale with weight two,
whereas the complex structures $J^A{}_{\!B}$ are invariant.

Secondly, one can verify that the 
isometries introduced in subsection~2.2 commute with scale
transformations, provided that  
\be
k_I^A = \kild{B}\,D_B k^A_I\,.  \label{killing-scale}
\ee
This leads to another identity, 
\be
g_{AB} \, k_I^A\, \kild{B} = 0\,. \label{killing-ortho}
\ee
In particular these results hold for the SU(2) Killing vectors and
imply, in addition, that the latter commute with the
tri-holomorphic isometries. To see this, one writes
$k_{ij}^B\,D_Bk_{IA}$ as $\kild{B} \,D_A\pa_B P_{Iij}$ using
\eqn{su(2)-vector}, \eqn{moment} and the fact that the complex
structures are covariantly constant. Interchanging the order of the
derivatives and extracting the complex structure then gives 
\be
k^B_{ij} \,D_Bk_I^A = J^A_{ijB}\, k_I^B\,,
\ee
which implies that the tri-holomorphic Killing vectors commute with
SU(2). From the above equations one can derive the following
result for the variation of the moment maps under a dilatation,
\be
\kild{A}\,\pa_A P_I^{ij} = J^{ij}_{AB} \,\kild{A}\,
k^B_I=-k^{ij}_Ak^A_I = 2  P_I^{ij} \,, \label{scaling-P}
\ee
i.e. they scale with conformal weight 2. Here we have adjusted an
integration constant in $P_I^{ij}$ in the last equation.  Combining
the above equation with previous results, one establishes that the moment
maps transform under SU(2) according to 
\be
k^A_{kl}\, \pa_A 
P_I^{ij}= 2\delta^{(i}_{(k}\varepsilon_{l)m}\, P^{j)m}_I\ .
\ee
The latter expresses the fact that the moment maps form a triplet
under SU(2). It is then easy to check that the action is invariant
under dilatations, U(1) and SU(2).
\subsection{Hyper-K\"ahler potential and ${\rm Sp}(n)\times {\rm Sp(1)}$
sections} 
The existence of the homothetic Killing vector satisfying
\eqn{homo-vector} has important consequences for the geometry. First
of all \eqn{homo-vector} implies that 
$\kild{A}$ can (locally) be expressed in terms of a potential 
$\chi$, according to 
$k_{{\scriptscriptstyle\rm D}\,A} 
=\pa_A\chi$. Up to a suitable additive 
integration constant, one can then show that \cite{GibbonsRych}
\be
\chi(\phi) = \ft12 g_{AB}\,\kild{A}\,\kild{B}\,. \label{HKpot}
\ee
Observe that $\chi$ is positive for a space of positive signature. A
second (covariant) derivative acting on  $\chi$ yields the metric, and
therefore a third derivative vanishes, 
\be
D_AD_B\chi = g_{AB}\;,\qquad
D_AD_BD_C\chi = 0\;. \label{derivativechi}
\ee
The first condition expresses the fact that the metric is the second
(covariant) derivative of some function, somewhat analogous to the
\Ka\ potential in \Ka\ metrics, but now written in real
coordinates. A \Ka\ potential is guaranteed to exist for any
hyper-K\"ahler space, but the potential
$\chi$ does not always exist. In the
literature $\chi$ is sometimes called the
hyper-K\"ahler potential (see, 
e.g. \cite{Swann,Galicki}). This means that $\chi$ 
serves as a \Ka\ potential for each of the three complex
structures, as follows from the following equation, 
\be
\ft12(\d_A^{\,C} + J^{\Lambda}{}_{\!\!A}{}^C )\, (\d_B^{\,D} -
J^{\Lambda}{}_{\!\!B}{}^D )\, 
D_CD_D \chi = J^{\Lambda}_{AB} \,, 
\ee
where $J^\Lambda= (\sigma_2\sigma^\Lambda)_{ij} \,J^{ij}$ and
$\Lambda=1,2,3$ is kept fixed.

The hyper-\Ka\ potential $\chi$ is invariant under
isometries, as follows directly from \eqn{killing-ortho}. In
particular it is invariant under the SU(2) isometry; explicitly, 
\be
\d \chi = 
(\theta_{\scriptscriptstyle\rm SU(2)})^{i}{}_k\varepsilon^{jk} \,
k_{ij}^B 
\,\partial_B   \chi = 
(\theta_{\scriptscriptstyle\rm SU(2)})^{i}{}_k\varepsilon^{jk}\,
J_{ij\,AB}\,k^B _{\scriptscriptstyle\rm D} k^A_{\scriptscriptstyle\rm
D} = 0\,,
\ee
where we made use of \eqn{su(2)-vector}. However, it is not invariant
under dilatations,
\be
\d\chi =\kild{B} \,\pa_B\chi =2 \chi\,. 
\ee

Another interesting consequence of the homothety is that it enables a
reformulation in terms of local sections of an Sp$(n)\times{\rm
Sp}(1)$ bundle. The existence of such a so-called associated
quaternionic bundle is known from general arguments \cite{Swann}.
These sections are defined from the $S$-supersymmetry variation of 
the hypermultiplet spinors (c.f. \eqn{eq:dSTwo}), 
\be \label{sections}
A_i{}^\a (\phi) \equiv \kild{B}(\phi)\,V_{B\,i}^\a(\phi)\,.
\ee
They satisfy a quaternionic pseudo-reality condition
\be
A^{i\,\bar \a} \equiv (A_i{}^\a)^\ast = \varepsilon^{ij}\,
\bar\Omega^{\bar\a\bar\b}\, G_{\bar \b\gamma} \, A_j{}^\gamma\,,
\label{realquat}
\ee
as follows from \eqn{pseudoreal}.  
Using \eqn{homo-vector} one proves that the covariant
derivative of $A_i{}^\a$ reproduces the quaternionic vielbeine,
\be
D_B A_i{}^\a = V_{B\,i} ^\a\,, \qquad \varepsilon_{ij}\,\Omega_{\bar
\a\bar\b}\, D_B A^{j\bar\b} = g_{BC} \, \gamma_{i\bar\a}^C \,.
\label{DA=V} 
\ee
One easily verifies that the hyper-\Ka\ potential
$\chi$ can be written as 
\bea
\chi = \ft12\, g_{AB}
\,k_{\scriptscriptstyle\rm D}^A\, k_{\scriptscriptstyle\rm D}^B
= \ft12 G_{\bar \b \a}\,A_i{}^\a A^{i\bar \b} =\ft12  \varepsilon^{ij}
\bar\Omega_{\a\b}\,A_i{}^\a A_j{}^\b\ ,  \label{surface}
\eea
or
\be
\bar\Omega_{\a\b}\,A_i{}^\a A_j{}^\b = \varepsilon_{ij}
\,\chi \,. \label{AA-id} 
\ee
We also note the following identity, 
\be
J^{ij}{}_{\!B}{}^C \,D_C A_k{}^\a = - \d^{(i}_k \varepsilon ^{j)l}\, D_B
A_l{}^\a \,. \label{JDA-id}
\ee
Furthermore we have 
\begin{equation}
{R_{AB}}^\a{}_{\!\b} \,A_i{}^\b={R_{AB}}^\a{}_{\!\b}\, \bar\Omega_{\a\g} \,
A_i{}^\g= 0\ , \label{spn-1}
\end{equation}
which is a consequence of $D_AD_BA_i{}^\a=0$ and the symplectic nature
of the curvature ${R_{AB}}^\a{}_\b$.  This implies that the generic
holonomy group is now reduced from Sp($n$) to Sp($n-1$). 
Also, using \eqn{su(2)-vector}, \eqn{AA-id} and \eqn{JDA-id}, one
finds  
\be
\kild{B} \, D_BA_i{}^\a = A_i{}^\a\,,\qquad 
k^{ij\,B}\, D_B A_k{}^\a =  \d_k^{(i}\varepsilon^{j)l}\, A_l{}^\a \,,
\ee
so that 
\be
\bar\Omega_{\a\b} \,A_i{}^\a \,D_B A_j{}^\b = \ft12 \varepsilon_{ij}
\,\kild{}_B + k_{ij\,B}  \,. \label{ADA-id}
\ee
Applying a second derivative $D_A$ to the above relation gives
\be
\bar\Omega_{\a\b} \,D_AA_i{}^\a \,D_B A_j{}^\b = \ft12 \varepsilon_{ij}
\,g_{AB} - J_{ij\,AB}  \,.\label{g-J}
\ee
Note that the quantities in \eqn{g-J} have weight 2 under the
homothety. 
For future use we also recall some earlier results, but now expressed in
terms of the local sections,
\bea
g^{AB} \,D_A A_i{}^\a\,  D_B A_j{}^\b  &=& \varepsilon_{ij} \,
\Omega^{\a\b}\,,    \nonumber \\
 g^{AB} \,D_A A_i{}^\a\, D_B A^{j\bar\b} &=& \d^j_i \,
G^{\a\bar\b}\,,  \nonumber \\
R_{AB}{}^{\!\g}{}_{\!\a}\,\bar \Omega_{\g\b} \,
D_CA_i{}^\a\, D_DA_j{}^\b \, \varepsilon^{ij} &=& R_{ABCD}\,.
\label{new-old-results} 
\eea

\subsection{The hypermultiplet action and transformation rules}

In this subsection, we write the hypermultiplet action and 
transformation  rules in terms of the sections $A_i{}^\a(\phi)$ introduced in
\eqn{sections}. 
The complete Lagrangian, including the terms associated with gauged
isometries, can be written as 
\begin{eqnarray}
{\cal L}&=&-\ft12G_{\bar \a \b}\,
{\cal D}_\mu A_i{}^{\b} \,{\cal D}^\mu A^{i\,\bar \a} 
-G_{\bar \a \b}( \bar\zeta^{\bar \a} {\cal D}\!\slash \,\zeta^\b +  
\bar\zeta^\b {\cal D}\!\slash \,\zeta^{\bar\a}) -\ft14  W_{\bar
\a\b\bar\g\d}\, \bar \zeta^{\bar \a}  
\g_\m\zeta^{\b}\,\bar \zeta^{\bar \g} \g^\m\zeta^\d\nonumber\\
&&+\Bigl[\,2g\,{\bar X}^\g{}_\a\,{\bar \O}_{\b \g}\,{\bar \z}^\a\z^\b+2g\,
{\bar \z}^\a\,{\bar \O}_{\a\b}\,\Omega^{i\b}{}_\gamma\,A_i{}^\gamma
+\mbox{ h.c.}\Bigr]\nonumber\\
&&+2g^2\,G_{\bar \a \b}\,A^{i\bar \a}\,\bar X^\b{}_\g\,X^\g{}_\d 
\,A_i{}^\delta +\ft12 g\,A_i{}^\a\,{\bar \O}_{\a\b}\,Y^{ij\b}{}_\g\,
A_j{}^\g\,,  
\label{act-section}
\end{eqnarray}
where the covariant derivatives are defined by
\bea
{\cal D}_\mu A_i{}^\a &=&\partial_\mu A_i{}^\a +\pa_\mu\phi^A\,
{\G_A}^{\!\a}{}_{\!\b}  \,
A_i{}^\b -g\,W_\m{}^{\!\!\a}{}_{\!\!\b}\,A_i{}^\b\, , \nonumber\\
{\cal D}_\mu\z^\a &=&\partial_\mu \z^\a+\pa_\mu\phi^A\,
{\G_A}^{\!\a}{}_{\!\b} \, 
\z^\b -g\, W_\m{}^{\!\!\a}{}_{\!\!\b}\,\z^\b\, , \; \label{cov-der}
\eea
and we have used Lie-algebra valued vector multiplet fields associated
with gauged isometries, 
$W_\m{}^{\!\!\a}{}_{\!\!\b}$, $X^\a{}_\b$, $Y^{ij\a}{}_\b$ and
$\Omega^{i\a}{}_\b$ (for
the precise definition, see below),  
In addition to the equation in the previous subsection we made use of
the identities, 
\bea
k^A_I\,V_{Ai}^\a&=& k^A_I\, D_A A_i{}^\a = {t_I}^\a{}_{\!\b}\, A_i{}^\b\,,
\nonumber \\ 
P_{I\,ij}&=& -\ft12\, k_{A\,ij}\, k^A_I = - \ft12 \bar\Omega_{\a\b}\, A_i{}^\a
\,(t_I)^\b{}_\g\,A_j{}^\g \ .
\eea
The first relation follows from \eqn{t-V-relation} and
\eqn{killing-scale},  and for the second equation we made use of the
last equality in \eqn{scaling-P}.

The action may be compared to the one in \cite{DWLVP} (more precisely, to
the part that pertains to the rigidly supersymmetric Lagrangian).
However, in that reference, 
the $A_i{}^\a$  are identical to the coordinate fields, whereas in the
present more general case they are local sections as
explained in subsect.~3.2. Because the target-space manifold is not
flat, we encounter a nontrivial metric in \eqn{act-section} as well as
nontrivial Sp($n$) connections in the
covariant derivatives \eqn{cov-der}. Furthermore, the generators 
$t_I(\phi)$ associated with the 
isometries are not constant, but depend on the scalar fields as we
indicated before. This means that the Lie-algebra valued vector
multiplet fields associated with the gauged isometries depend also on
the hypermultiplet scalars. Their definitions are
\bea
W_\m{}^{\!\!\a}{}_{\!\!\b}&=& W_\m^I\,[t_I(\phi)]^\a{}_\b \,,
\nonumber \\
X^\a{}_\b &=& X^I\,[t_I(\phi)]^\a{}_\b\,,\qquad \bar X^\a{}_\b =
\bar X^I\, [t_I(\phi)]^\a{}_\b\,, \nonumber \\
Y^{ij\a}{}_\b &=& Y^{Iij} \,[t_I(\phi)]^\a{}_\b\,, \nonumber \\
\Omega^{i\a}{}_\b &=& \Omega^{I\,i}\,[t_I(\phi)]^\a{}_\b\,,\qquad
\Omega_i{}^\a{}_\b 
= \Omega^I_i\,[t_I(\phi)]^\a{}_\b\,.
\eea
Nevertheless, the correspondence with the
formulation in \cite{DWLVP} will be helpful later on when evaluating the
coupling to conformal supergravity. 

In order to obtain the transformation rules of the Sp$(n)\times{\rm
Sp}(1)$ sections under dilations, SU(2) and isometry transformations,
we use the general relation 
\be
\d A_i{}^\a = \d\phi^B \pa_B A_i{}^\a= \d\phi^B \,
V_{B\,i}^\a - \d \phi^B \G_B{}^{\!\a}{}_{\!\b} A_i{}^\b \,.
\ee
Using \eqn{bilinear-id}, \eqn{t-V-relation} and 
\eqn{killing-scale}, we then find for a combined dilatation, 
chiral transformation and target-space isometry, that
\be
\d A_i{}^\a =  \theta_{\scriptscriptstyle\rm D} A_i{}^\a +
 (\theta_{\scriptscriptstyle \rm SU(2)})_i{}^j \, A_j{}^\a  + 
g\theta^I \, 
t_I{}^\a{}_{\!\b}\,  A_i{}^\b  - \d \phi^A \G_A{}^\a{}_{\!\b} 
A_i{}^\b  \,.
\ee
This result should be combined with that for the fermions,  
derived in the previous section,
\be
\d\zeta^\a =  \ft32 \theta_{\scriptscriptstyle\rm D}\, \zeta^a 
-\ft12 i\theta_{\scriptscriptstyle\rm U(1)} \,\zeta^\a  + g\theta^I \,
t_I{}^\a{}_{\!\b}\,  \zeta^\b  - \d \phi^A \G_A{}^\a{}_{\!\b} 
\zeta^\b  \,. 
\ee
Similarly we determine the transformations under 
$Q$- and $S$-supersymmetry,
\bea
\d A_i{}^\a &=&  2\, \bar\e_i\zeta^\a + 2\, 
\varepsilon_{ij} \,
G^{\a\bar\b}\Omega_{\bar\b\bar\gamma} \,
\bar\e^j\zeta^{\bar\gamma}   - \d_{\scriptscriptstyle\rm Q}  
\phi^B \G_B{}^\a{}_{\!\b} \,A_i{}^\b \,,\nonumber \\
\d \zeta^\a &=& {\cal D}\!\slash \,A_i{}^\a \e^i- 
\d_{\scriptscriptstyle\rm Q} \phi^B \G_B{}^{\!\a}{}_{\!\b}\, \zeta^\b 
 + 2g\, X^{\a}{}_{\!\b}\,A_i{}^\b \,
\varepsilon^{ij}\e_j  + A_i{}^\a\,\eta^i\,,\nonumber\\ 
\d \zeta^{\bar \a}  &=& {\cal D}\!\slash\, 
A^{i\,\bar \a} \e_i- \d_{\scriptscriptstyle\rm Q}  
\phi^B \G_B{}^{\!\bar\a}{}_{\!\bar\b}\,\zeta^{\bar\b}  +2g\,\bar 
X^{\bar\a}{}_{\!\bar \b}\, A^{i\,\bar\b}\,
\varepsilon_{ij}\e^j +  A^{i\, \bar \a}\,\eta_i  \,.\label{QS-transf}
\eea

Again, we stress that, apart from the Sp$(n)$ connection (and a 
slight change in notation), these transformation rules are identical
to the ones specified for a flat target space \cite{DWLVP}, where the
local sections can be identified directly with the target-space coordinates. 

Finally, we recall that it is straightforward to write down actions for the 
vector multiplets that are invariant under rigid $N=2$ 
superconformal transformations. Those are based on a holomorphic 
function that is homogeneous of degree two \cite{DWVP}.

\section{Cone structure and quaternionic geometry}
\setcounter{equation}{0}
In this section we discuss the properties of the special
hyper-K\"ahler space.  We will show how this space can be described as a cone
over a tri-Sasakian manifold. The latter spaces (which are of
dimension $4n-1$) are characterized by the existence of three $(1,1)$
tensors and three Killing vectors that are subject to certain
conditions. A manifold is
tri-Sasakian if and only if its cone is hyper-K\"ahler. Tri-Sasakian
spaces are Einstein and take the form of an Sp(1) fibration over a 
quaternionic space.  This quaternionic space is the one that appears
in the coupling of hypermultiplets to supergravity (for more details,
see \cite{sasaki}, where the relation 
between special hyper-\Ka, tri-Sasakian and quaternionic spaces is
reviewed from a more mathematical viewpoint).  

We start by noting that the Riemann tensor vanishes upon
contraction with any one of the four vectors $(\kild{A}, k_{ij}^A)$,
i.e. 
\be
R_{ABCE} \,\kild{E} = 0\, ,\qquad R_{ABCE} \,k_{ij}^E  = 0\ . 
\label{Rhorizontal}
\ee
The first equation \eqn{Rhorizontal} is derived by antisymmetrizing
the second equation \eqn{derivativechi} in the indices $[AB]$. The  
second equation \eqn{Rhorizontal} follows from inserting 
\eqn{D-su(2)} into \eqn{symmetric}. 
Incidentally, \eqn{Rhorizontal} implies that the Ricci tensor has at 
least four null vectors. However, in the case at hand this poses no
extra restrictions as hyper-K\"ahler spaces are Ricci-flat. The
above results can also derived from the fact that the Sp($n$) holonomy
group is reduced to Sp($n-1$), c.f. \eqn{spn-1}.  This follows from
applying \eqn{ADA-id}. 

We recall that these four vectors are orthogonal (cf. \eqn{su(2)-vector},
\eqn{HKpot}), 
\be
\kild{A} \, \kild{}_A = 2\,\chi\,, \qquad
k^{A}_{ij} \,k_A^{kl}= \d_{(i}^k\d_{j)}^l 
\,\chi\,,\qquad    \kild{A} \,k_A^{ij} = 0\,.
\ee
This implies that the hyper-K\"ahler manifold is locally a product
${\bf R}^4 \times {\bf Q}^{4n-4}$, where ${\bf R}^4$ denotes a flat
four-dimensional space. By 
decomposing  ${\bf R}^4$ as $R^+\times S^3$, we can write the
hyper-K\"ahler manifold as a cone over a so-called tri-Sasakian manifold;
the latter is then a fibration of Sp(1) over ${\bf Q}^{4n-4}$. Hence
the manifold can be written as\footnote{
Strictly speaking it is Sp$(1)/{\bf Z_2}$ where Sp(1) is the group
that acts on the quaternionic vielbeine and on the sections introduced in
the previous chapter.} 
$R^+ \times[{\rm Sp(1)} \times {\bf Q}^{4n-4}]$. Spaces with a
homothety can always be described as a 
cone. This becomes manifest when decomposing the coordinates $\phi^A$
into coordinates tangential and orthogonal to the $(4n-1)$-dimensional
hypersurface defined by setting $\chi$ to a constant. The line element
can then be written in the form \cite{GibbonsRych},  
\be
ds^2 = {d\chi^2\over 2 \chi} + 2 \chi\,
h_{ab}(x) \,dx^a\,dx^b \,, \label{conemetr}
\ee
where the $x^a$ are the coordinates associated with the
hypersurface\footnote{%
  In terms of a radial variable $r^2=2\chi$, this yields the 
  usual form of a cone metric
  \[
  ds^2=dr^2+r^2\,h_{ab}(x)\,dx^adx^b\ .  
  \] 
}. %
In the present case this hypersurface must be a tri-Sasakian space and
the hyper-K\"ahler space is therefore a cone over the tri-Sasakian space. 

The purpose of the remainder of this section is to establish that
${\bf Q}^{4n-4}$ is a quaternionic manifold. In the next section we
show how ${\bf Q}^{4n-4}$ arises in the coupling of hypermultiplets to
supergravity. The tangent space of the hyper-\Ka\ space
can be decomposed into the four directions along $(\kild{A}, k_{ij}^A)$,
and a $(4n-4)$-dimensional space ${\bf Q}^{4n-4}$ that is locally
orthogonal to that. Tensors that vanish upon contraction with
$(\kild{A}, k_{ij}^A)$ will be called {\it horizontal}. 

Let us  introduce a vector field ${\cal V}_{A\,ij}$ which will serve
as a connection for Sp(1) in a way that will become clear shortly,
\be 
{\cal V}_{A\,ij} = {k_{ij\,A}\over  \chi}=
J_{ij\,A}{}^B\, \pa_B \, \ln \chi\;.
\label{sp1conn}
\ee
This vector field is invariant under target-space dilatations
and gauge isometries, i.e.
\bea
\d_{\scriptscriptstyle \rm D} {\cal V}_{A\,ij}&=&\kild{B}\,\partial_B 
{\cal V}_{A\,ij}+\partial_A \kild{B}\,  {\cal
V}_{B\,ij}=0\,,\nonumber\\ 
\d_{\scriptscriptstyle \rm G} {\cal V}_{A\,ij}&=&k_I^B\,\partial_B
{\cal V}_{A\,ij}
+\partial_A k_I^B\,  {\cal V}_{B\,ij}=0\ ,
\eea
and rotates under target-space SU(2), as follows from 
\begin{equation}
\d {\cal V}_A^{ij}= 
k^{Bkl}\,\partial _B{\cal V}_A^{ij}+\partial_A k^{Bkl} \,{\cal V}_B^{ij}=
2\varepsilon^{(i(k}\,{\cal V}_A^{l)j)}\ .
\end{equation}
With ${\cal V}_{A\,ij}$ we associate an Sp(1) curvature tensor, 
\begin{eqnarray}
R_{AB\,ij} &\equiv&  \pa_A {\cal V}_{B\,ij} - \pa_B {\cal V}_{A\,ij}
- \varepsilon^{kl} \Big({\cal V}_{A\,ik}\,{\cal V}_{B\,jl} + {\cal
V}_{A\,jk}\,{\cal V}_{B\,il}\Big)\nonumber\\
&=& \chi^{-1} \Delta_{\a\b}\,\Big[ D_A A_i{}^\a\, D_B A_j{}^\b + D_A
A_j{}^\a \, D_B A_i{}^\b \Big] \,, \label{sp1-curv}
\end{eqnarray}
where we have used the definition 
\begin{equation}
\Delta_{ \a\b}= {\bar
\Omega}_{\a \b} -\frac{1}{\chi } \,\varepsilon^{kl} 
({\bar \Omega}_{\a \gamma }A_k{}^\gamma)\,({\bar \Omega}_{\b \delta}
A_l{}^\delta)\,.   \label{def-Delta}  
\end{equation}
Observe that $\Delta_{\a\b}$ is a projection operator, i.e., it
satisfies $\Delta_{\a\b} \,\Omega^{\b\gamma}\Delta_{\gamma\d}= -
\Delta_{\a\d}$, and 
it projects onto the $(2n-2)$-dimensional subspace orthogonal to the
$A_i{}^\a$,  
\begin{equation}
\Delta _{\a \b}\, A_i{}^\b = 0\ .
\end{equation}
Note that we have $\kild{B}\,D_B \Delta_{\a\b}= k^{B}_{ij} \,D_B
\Delta_{\a\b} =0$, so that $\Delta_{\a\b}$
is invariant under dilatations and SU(2) transformations. One
can also show that $\Delta_{\a\b}\,D_B A_i{}^\b$ is horizontal, i.e.,
\be
\kild{B} \,\Delta_{\a\b}\,D_BA_i{}^\b=
k^B_{ij}\,\Delta_{\a\b}\,D_BA_i{}^\b = 0\,. 
\ee

The identity \eqn{sp1-curv} can be generalized to
\be
\chi^{-1} \Delta_{\a\b} D_AA_i{}^\a\, D_B A_j{}^\b = \ft12
\varepsilon_{ij} \, G_{AB} + \ft12 R_{AB\,ij}\,, \label{G-R}
\ee
where
\be
G_{AB} = \chi^{-1} \varepsilon^{ij}\,\Delta_{\a\b} D_AA_i{}^\a\, 
D_B A_j{}^\b \,.
\ee
Observe that both $G_{AB}$ and $R_{ABij}$ are of zero weight under the
homothety and  are horizontal,
i.e., they vanish upon contraction with any of the four vectors $(\kild{A},
k_{ij}^A)$, and are thus orthogonal to the corresponding (local)
four-dimensional subspace. 

The tensor $G_{AB}$ will provide a  metric for ${\bf
Q}^{4n-4}$. The relation between $G_{AB}$ and the hyper-\Ka\ metric
$g_{AB}$ is given by 
\begin{eqnarray}
g_{AB}&=&\frac{1}{2\chi}\,\kild{}_A\,\kild{}_B
+\frac{1}{\chi}\,k_{A\,ij}\,k_{B}^{ij}
+\chi \, G_{AB}\nonumber\\
&=&\frac{1}{2\chi}\,\kild{}_A\,\kild{}_B
+\chi\Big[{\cal V}_{A\,ij}\,{\cal V}_{B}^{ij}+G_{AB}\Big]\ , \label{def-G}
\end{eqnarray}
where we have used \eqn{ADA-id} and \eqn{g-J}. Observe that this
relation reflects both the cone structure of the hyper-K\"ahler space
and the Sp(1) fibration of the tri-Sasakian space. 
It is not possible to give an explicit expression for the inverse
metric, at least not in general, but this is not really needed in view
of the horizontality of $G_{AB}$. When acting on horizontal tensors, 
$\chi\, g^{AB}$ acts as the inverse metric in view of the
identity 
\be
G_{AC}\,g^{CD}\,G_{DB} = \chi^{-1} G_{AB}\,.
\ee
We already showed that $\Delta_{\a\b}\,D_BA_i{}^\b$ was horizontal,
and conversely, the horizontal projection $G_{AB} \, g^{BC}
\,D_CA_i{}^\a$ is in 
the $(2n-2)$-dimensional eigenspace projected onto by
$\Delta_{\a\b}$. Therefore 
$\Delta_{\a\b}\,D_BA_i{}^\b$ is a candidate for the quaternionic
vielbein associated with ${\bf Q}^{4n-4}$ and $\Delta_{\a\b}$ projects
onto the tangent space of ${\bf Q}^{4n-4}$. More precisely, we
introduce the following related sets of $4n-4$ vectors,  
\bea
\hat V_{Ai}^\a &\equiv& - {1\over \sqrt\chi} \, \Omega^{\a\b}
\,\Delta_{\b\g} \,V_{Ai}^\g =  - {1\over \sqrt\chi} \, \Omega^{\a\b}
\,\Delta_{\b\g} \,D_A A_{i}{}^\g  \,,\nonumber\\
\hat\g_{Ai\bar\a} &\equiv& {1\over \sqrt\chi} \, G_{AB}
\,\g^B_{i\bar\a} = 
\varepsilon_{ij} 
\,\bar\Delta_{\bar\a\bar\b} \, \bar{\hat V}{}_{\!A}^{j\bar\b}\,, 
\label{def-reduced-vielbeine}
\eea
which satisfy algebraic relations that are completely analogous to
those satisfied by the quaternionic vielbeine of the
hyper-K\"ahler space. In particular we note that $\hat V_A$ and
$\hat\gamma_A$ are each other's inverse in the reduced
$(4n-4)$-dimensional space,
\be
\chi\,g^{AB}\, \bar{\hat \gamma}{}_{\!A\a}^i\, \hat V_{Bj}^\b   =
\d^i_{\,j} \, \Delta_{\a\g}\,\Omega^{\b\g}\,,\qquad 
\varepsilon^{ij} \,\Delta_{\a\b}\, \hat V_{Ai}^\a\,\hat V_{Bj}^\b =
G_{AB} \,, 
\ee
where $\Delta_{\a\g}\,\Omega^{\b\g}$ is the identity matrix projected
onto the $(2n-2)$-dimensional subspace.  
The significance of these results will become clear in due course.

Subsequently we note that
there exists an identity similar to \eqn{def-G} which relates the
complex structures to the field strength $R_{ABij}$,
\be
J_{ij\,AB} = - {1\over \chi} \,\Big[\kild{}_{[A}\, k_{ijB]}
+ \varepsilon^{kl}\, k_{ki[A}\,k_{lj B]}\Big] - \ft12 \chi\, R_{ABij}
\, .  \label{J-id}
\ee
This motivates us to introduce the following tensors, 
\begin{equation}
{\cal J}_{AB}^{ij}=J^{ij}_A{}^CG_{CB}\ .
\end{equation}
A straightforward calculation using \eqn{J-id} shows
that they satisfy  
\begin{equation}
{\cal J}_{ABij}=-\ft12 R_{ABij}\ , \label{sp1-hol}
\end{equation}
so that the ${\cal J}_{ABij}$ are antisymmetric, horizontal and scale
invariant. Furthermore these tensors satisfy the product rule 
\begin{equation}
\chi\, {\cal J}^{ij}_{AC}\,g^{CD}\,{\cal J}^{kl}_{DB}=\ft12
\varepsilon^{i(k}\varepsilon^{l)j}\, G_{AB}
+ \varepsilon_{~}^{(i(k} \, {\cal J}^{l)j)}_{AB}\ ,
\label{calJcalJ}
\end{equation}
which is similar to \eqn{JJ}. The tensors ${\cal J}_{ABij}$
are candidate almost-complex structures in the horizontal subspace ${\bf
Q}^{4n-4}$. Under SU(2) target-space  
transformations they rotate into each other according to 
\begin{equation}
k^C_{kl} \,\pa_C {\cal J}_{ABij} + \pa_A k^C_{kl} \,{\cal J}_{CBij}
+\pa_B k^C_{kl} \,{\cal J}_{ACij}   = 2\,\varepsilon_{(i(k} \,{\cal
J}_{AB\,l)j)}\, . \label{su2-acc}
\end{equation}

Given a horizontal tensor $H_{AB\cdots}$ that is
invariant under the 
homothety and the SU(2) target-space transformations, then the
covariant derivative of such a tensor is no longer horizontal. This can
be cured by making use of a modified covariant derivative $\hat D_A$,
defined so that the following properties hold,
\begin{eqnarray}
\kild{A}\,{\hat D}_C\,H_{AB\cdots}=\kild{C}\,{\hat D}_C\,H_{AB\cdots}=0 \
,\nonumber\\ 
k^A_{ij}\,{\hat D}_C\,H_{AB\cdots}=k^C_{ij}\,{\hat D}_C\,H_{AB\cdots}=0
\ .
\end{eqnarray}
The modified derviative is obtained by using a modified target-space
connection, 
\begin{equation}
\hat\G_{AB}{}^C= \G_{AB}{}^C -   \d_{(A}^C\, \pa_{B)}  \ln
\chi +2 {\cal V}_{(A\,ij} \,J_{\,B)}^{ij\,C} \ .  \label{Q-conn}
\end{equation}
Because the modification is symmetric in $(A,B)$, the connection
remains torsion free. Observe that $\hat D_A (\chi g^{BC})$, $\hat
D_A\kild{B}$ and $\hat D_Ak_{ij}^B$ should be zero when contracted
with a horizontal tensor. This 
is obviously the case as can be seen from the formulae,
\bea
\hat D_A (\chi g^{BC}) &=& - \d_A^{(B}\,\kild{C)} + 2 J_{\,A}^{ij (B}
\, k^{C)}_{ij} \,, \nonumber \\
\hat D_A\kild{B} &=& \chi^{-1} \Big(-\ft12 \kild{}_A\,\kild{B} +
k_{ij \,A}\,k^{ij\,B}\Big)\,,\nonumber \\
\hat D_A k_{ij}^B &=&\chi^{-1}\Big( - \ft12 \kild{}_A\,k^B_{ij} +
\ft12 k_{ij\,A}\,\kild{B} -  k^{kl}_A\, \varepsilon_{k(i} \,k_{j)l}^B\Big) \,. 
\eea
The above construction can be generalized to tensors $H$ that carry also
SU(2) indices, indicating that they 
transform covariantly under target-space SU(2) transformations,
e.g. as in $k_{kl}^A\, \pa_A H^i = \d^i_{(k}\varepsilon_{l)j}\, H^j$
in the simplest case. Then
one can show that the derivatives of these tensors are still
horizontal, provided  one covariantizes
$\hat D_A$ and includes an SU(2) connection ${\cal V}_{A\,ij}$. The
crucial identity for showing this is $k^A_{ij}\, {\cal V}^{kl}_A =
\d^k_{(i}\, \d^l_{j)}$.

With respect to the new connection, $G_{AB}$ is covariantly constant,
\begin{equation}
{\hat D}_C\,G_{AB}=0\ , \label{metric-G}
\end{equation}
so that the new connection must be just the Christoffel connection
associated with $G_{AB}$. Likewise the tensors ${\cal J}_{ABij}$ are
covariantly constant modulo a rotation that involves the Sp(1) 
connection, 
\begin{equation}
{\hat D}_C {\cal J}_{ABij}=2\,{\cal V}_{Ck(i} \,{\cal
J}_{ABj)l}\,\varepsilon^{kl} \,. 
\end{equation}
Note that the terms on the right-hand side covariantize the
derivative on the left-hand side with respect to SU(2). 
Hence the tensors ${\cal J}_{ABij}$ define three almost-complex structures in
the space ${\bf Q}^{4n-4}$ which are covariantly constant up to an
Sp(1) rotation proportional to the Sp(1) connections. This implies
that ${\bf Q}^{4n-4}$ is a quaternionic space (see e.g. \cite{KNY}).  

To verify this result, let us compute the Riemann tensor associated
with the new connection \eqn{Q-conn}. 
\be
{\hat R}_{ABC}{}^{\!D} = R_{ABC}{}^{\!D} -  G_{C[A} \,\delta^D_{B]}
+ R_{ABij}\,J^{ijD}_{\,C}  
- R_{C[Aij}\, J^{ijD}_{\,B]}\ .
\ee
Observe that the right-hand side is not horizontal, but by
construction (via the Ricci identity) is horizontal when acting on
a horizontal tensor with lower index $D$. Hence, when lowering the
index by contraction with the metric $G_{DE}$ one must obtain a
horizontal tensor. This is confirmed by explicit construction,
\bea
{\hat R}_{ABCD} &\equiv& {\hat R}_{ABC}{}^{\!E} \,G_{ED} \nonumber\\
&=& \chi^{-1} \,R_{ABCD} +G_{D[A}\,G_{B]C}  
+ R_{ABij}\,{\cal J}^{ij}_{CD}  
- R_{C[Aij}\, {\cal J}^{ij}_{B]D}\ .
\eea
By virtue of \eqn{sp1-hol} $\hat R_{ABCD}$ has all the symmetry
properties of a Riemann tensor. Observe that the explicit factor of
$\chi^{-1}$ arises because the original curvature of the
hyper-K\"ahler manifold is defined by lowering the upper index by
means of the metric $g_{DE}$. Furthermore it satisfies the Bianchi
identity $\hat D_{[A}\hat R_{BC]DE}=0$.

Let us now calculate the Ricci tensor, which is symmetric by virtue of
\eqn{metric-G},  
\be
\hat R_{AB}= \chi\,\hat R_{ACBD}\, g^{CD} = -2(n+1) \,G_{AB}\,.
\ee
Oberve that we
used that the original hyper-K\"ahler manifold was Ricci flat and
that $G_{AB}\,g^{AB}\,\chi= 4(n-1)$.   
We may also verify the expressions for the Sp(1) holonomy 
\be
\hat R_{ABCD} \,g^{CE}g^{DF}\chi^2 {\cal J}^{ij}_{EF} = -4(n-1)\,
{\cal J}^{ij}_{AB}\,,
\ee
where we used that the original hyper-K\"ahler manifold has zero Sp(1)
holonomy. These are the expected results \cite{Ishihara} for a
$(4n-4)$-dimensional quaternionic manifold with Sp(1) curvature given
by \eqn{sp1-hol}. 

This completes the discussion of target-space properties. We now
return to aspects related to the Sp($n$) bundle over the special
hyper-\Ka\ space. First of all
we consider a modification of the connection $\G_A{}^{\!\a}{}_{\!\b}$
such that the the modified derivative of a tensor that is orthogonal
to $A_i{}^\a$ remains orthogonal. This requires that this derivative
acting on $A_i{}^\a$ must be proportional to $A_i{}^\a$ itself. When
combining this with a few other obvious requirements\footnote{
  In determining the precise modifications of the various connections,
  we were also guided to some extent by supersymmetry. However, 
  this aspect is postponed to sect.~5, where we outline the
  significance of the results of this section in the context of the
  coupling of   hypermultiplets to supergravity.
}, 
we arrive at the following connection,
\begin{equation}
{\hat \G}_A{}^{\!\a}{}_{\!\b} = {\G_A}^{\!\a}{}_{\!\b} - 
\frac{2}{\chi} \Big[ \varepsilon^{ij}\,   A_i{}^{\!(\a} \, D_A
A_j{}^{\g)} + A_i{}^{\a}\,A_j{}^\g \, {\cal V}_A^{ij}\, \Big]\,
\bar\Omega_{\g\b} \,. 
\label{new-conn}
\end{equation}
With this modification, the tensors $\bar \Omega_{\a\b}$ and
$G_{\bar\a\b}$  remain covariantly constant. The presence of the term
proportional to ${\cal 
V}_A^{ij}$  is required to preserve covariance with respect to
target-space SU(2) transformations.  This term also ensures that the
modification is horizontal. With the modified connection we
establish the required result, 
\be
\hat D_A A_i{}^\a = \ft12 \pa_A\ln\chi\, A_i{}^\a  + {\cal V}_{Aik} \,A_l{}^\a
\,\varepsilon^{kl}\,,  \label{A-cov-constant}
\ee
where the last term can be interpreted as an SU(2) covariantization of
the derivative on the left-hand side. 
The result \eqn{A-cov-constant} suffices to show that the modified
derivative of a tensor that is 
orthogonal to $A_i{}^\a$, will remain orthogonal. It is now obvious
that the projection operator 
$\Delta_{\a\b}$ is covariantly constant under the modified derivative 
\begin{equation}
{\hat D}_A  \Delta_{\a\b}=0\ .
\end{equation}

Including the modified connections $\hat\G_{AB}{}^C$ and 
$\hat\G_A{}^{\!\a}{}_{\!\b}$ as well as the SU(2) connection ${\cal
V}_A^{ij}$, one can explicitly verify that $\hat D_A V_{Bi}^\a$ is equal to
$\ft12 \pa_A\ln\chi\,V_{Bi}^\a$, up to terms that are proportional to
$A_k{}^\a$. This implies that the quaternionic vielbeine introduced
in \eqn{def-reduced-vielbeine} are covariantly constant with respect
to the new connections, so that we have
\be
\hat D_A (A_i{}^\a/\sqrt\chi) = \hat D_A \hat V_{Bi}^\a = \hat D_A
\hat \gamma_{B i\bar \a} = 0\,.
\ee
This result leads to two integrability relations
\bea
&&\hat R_{AB}{}^{\!\a}{}_{\!\b} \,A_i{}^\b - R_{ABik} \,
A_j{}^\a \,\varepsilon^{kj} = 0 \,,\nonumber\\
&& \hat R_{ABCD}\, \bar{\hat\gamma}{}^{Di}_\a +  \hat
R_{AB}{}^{\!\b}{}_{\!\a}\,\bar{\hat \g}{}^{i}_{C\b} + R_{AB}^{ik} \,
\bar{\hat\g}{}^{l}_{C\a} \,\varepsilon_{kl} = 0 \,. \label{integrab2}
\eea
Here $\hat R_{AB}{}^{\!\a}{}_{\!\b}$ is the curvature associated with
the new connection \eqn{new-conn}. We can explicitly evaluate this
tensor,
\be
{\hat R}_{AB}{}^{\!\a}{}_{\!\b} = R_{AB}{}^{\!\a}{}_{\!\b} +
 \frac{2}{\chi}\, \Omega^{\a\g}\Delta_{\g\d} \, \varepsilon^{ij} 
\,D_AA_i{}^{(\d} \, D_BA_j{}^{\e)} \, \Delta_{\e\b} - 
\frac{1}{\chi}\, 
R_{AB}^{ij}\, A_i{}^\a A_j{}^{\g}\, {\bar \Omega}_{\g \b}\ , 
\label{new-sympl-curv}
\ee
which indeed satisfies the first integrability relation. Note that all
expressions appearing in \eqn{new-sympl-curv} are horizontal.  

Now we recall that for a special hyper-K\"ahler manifold the tensor
$W_{\a\b\g\d}$ defined in \eqn{defW-symm} satisfies the constraint 
\be
W_{\a\b\g\d}\, A_i{}^\d = 0\,.
\ee
With this in mind we write the new curvature tensors as follows,
\bea
{\hat R}_{ABCD} 
&=& \ft12  \,\varepsilon^{ij} \,\varepsilon^{kl}\, \hat
V^\a_{Ai}\,\hat V^\b_{Bj}\,\hat V_{Ck}^\g\,\hat V_{Dl}^\d\; \hat
W_{\a\b\g\d} 
\nonumber \\
&& +G_{D[A}\,G_{B]C} - 2 {\cal J}^{ij}_{AB}\,{\cal J}_{CDij}  
+ 2{\cal J}^{ij}_{C[A}\, {\cal J}_{B]Dij}\,, \nonumber\\
\bar\Omega_{\a\e}\,{\hat R}_{AB}{}^{\!\e}{}_{\!\b} &=& -
\varepsilon^{ij} \,\hat V_{Ai}^\g\,\hat V_{Bj}^\d\,\Big[ \ft12\,
\hat W_{\a\b\g\d} +  2\Delta_{\a(\g}\,\Delta_{\d)\b} \Big] \nonumber\\
&& + \chi^{-1} 
R_{AB}^{ij}\, A_i{}^\g A_j{}^{\d}\,{\bar \Omega}_{\g \a}\, {\bar
\Omega}_{\d \b}\,,  \label{curvs-into-W}
\eea
where
\be
\hat W_{\a\b\g\d} \equiv \chi\,W_{\a\b\g\d}\,.
\ee
One can now verify that these curvatures satifsy also the second
integrability condition \eqn{integrab2}. We will return to this and
related issues in the next section. 

We close this section with a brief discussion of the isometries. For 
every tri-holomorphic Killing vector of the special hyper-\Ka\ manifold
we construct a corresponding vector in the horizontal manifold
${\bf Q}^{4n-4}$ by the projection
\be
\hat k_{I A} = G_{AB}\,k^B_I\,.
\ee
By explicit calculation one can then show that $\hat D_A \hat k_{IB} +
\hat D_B \hat k_{IA} =0$, so that we have a corresponding Killing
vector in the horizontal space and thus an isometry. Observe that the
SU(2) isometries of the special hyper-\Ka\ manifold 
do not generalize in this way, because the corresponding $\hat k_{IA}$
would simply vanish. This is not so surprising, as the SU(2)
acts on the corresponding tri-Sasakian space through its Sp(1) fibre. 

To study whether these isometries are tri-holomorphic in the
horizontal subspace, we first raise the index according to 
\begin{equation}
\hat k^A_I=\chi\, g^{AB} G_{BC} k^C_I=k^A_I+2 \, \hat
P_I^{ij}\,k^A_{ij}\, .  
\end{equation}
where ${\hat P}_{Iij}=\chi^{-1}P_{Iij}$.
The transformation of the almost complex structures in the horizontal
subspace is then governed by the expression, 
\begin{eqnarray}
\hat k^C_I\,\partial_C{\cal J}_{ABij}-2\partial_{[A}\hat k^C_I\,
{\cal J}_{B]Cij}&=&
k^C_I\,\partial_C{\cal J}_{ABij}-2\partial_{[A} k^C_I\, {\cal
J}_{B]Cij}\nonumber\\ 
&&+2{\hat P}_I^{kl}(k^C_{kl}\,\partial_C{\cal
J}_{ABij}-2\partial_{[A}k^C_{kl}\, 
{\cal J}_{B]Cij})\,, \;{~} \label{isom-acc}
\end{eqnarray}
The first line on the right-hand side is zero, as follows from
\eqn{J-id} and the fact that the isometries are 
tri-holomorphic and commute with dilatations and SU(2) in the special 
hyper-\Ka\ space. The second line is equal
to $4{\cal J}_{ABk(i}\,\varepsilon_{j)l}\, {\hat P}_I^{kl}$ by virtue
of \eqn{su2-acc}. We can 
now elevate the derivatives on the left-hand side to SU(2) covariant
dervatives. In this way we find
\be
\hat D_A({\cal J}_{BCij}{\hat k}^C_I)-\hat D_B({\cal J}_{ACij}\hat k^C_I)
=-2R_{ABk(i}\,\varepsilon_{j)l}\, \hat P_I^{kl}\,,
\ee
where we used the horizontallity of $\hat k^C_I$ and the Bianchi
identity for (or the covariant constancy of) $R_{ABij}\propto {\cal
J}_{ABij}$. The solution is given by  
\begin{equation}
{\cal J}_{ABij}\,\hat k^B_I=\hat D_A \hat P_{Iij}\, ,
\end{equation}
which can also be verified by explicit calculation. By substituting
previous results one verifies directly the modified 
equivariance condition, 
\begin{equation}
{\cal J}_{ABij}\,\hat k^A_I \hat k^B_J=-f_{IJ}{}^K\hat P_{Kij}
+4 \,\varepsilon^{kl}\,\hat P_{Ik(i} \,\hat P_{Jj)l}\ .
\end{equation}
The above results are in complete agreement
with the moment map construction for quaternionic manifolds
\cite{QKmoments,DFF}. The fact that the isometries generated by $\hat
k_I^A$ act consistently on horizontal tensors is ensured by the
following identities which follow from explicit calculation, 
\be
\kild{B}\, \hat D_B \hat k_I^A = k^{B}_{ij}\, \hat D_B \hat k_I^A =
0\,. 
\ee
Finally the algebra of the isometries is governed by
\be
\hat k^B_I\,\pa_B\hat k^A_J -\hat k^B_J\,\pa_B\hat k^A_I = - f_{IJ}{}^K\,
\hat k^A_K  + 2\,{\cal J}_{BCij}\,\hat k^B_I \hat k^C_J\, k^{Aij} \,.
\ee
Hence the algebra of isometries is satisfied up to SU(2). 

\section{Locally superconformal hypermultiplets}
\setcounter{equation}{0}

In this last section we consider the coupling of the hypermultiplets
to superconformal 
gravity.  To that order we introduce the Weyl multiplet, which
contains the gauge fields associated with the superconformal
symmetries as well as some extra matter fields 
\cite{DVV}. The bosonic gauge fields are  
the vielbeine $e_\m^a$, the spin-connection $\omega^{ab}_\m$, the
dilatational gauge field $b_\m$, the gauge field associated with special
conformal boosts $f_\m^{\,a}$ and the gauge fields associated with
SU($2)\times $U(1), denoted by ${V_\m}^{\!i}{}_j$ (antihermitean) and
$A_\m$. The fermionic gauge fields are the gravitino fields
$\psi^i_\m$ and the fields $\phi^i_\m$ associated with
$S$-supersymmetry. Finally, the matter fields are $T_{ab\,ij}$
(antisymmetric and selfdual in Lorentz indices and antisymmetric in
SU(2) indices), a spinor $\chi^i$ 
and a real scalar $D$. The fields $\omega_\m^{ab}$, $f_\m^a$ and
$\phi_\mu^i$ are  not independent and can be expressed in terms of the
other fields.  
We refer to \cite{DVV,DWLVP} for more details on the notation and
conventions.  
                                  
The transformation rules have been given in previous
sections, but will change in the context of local
supersymmetry. The most obvious change concerns the
replacement of the derivatives by derivatives that are covariant with
respect to the additional gauge symmetries.  
The derivatives covariant with respect to the bosonic gauge symmetries
for the scalar fields, the sections and the fermion fields, read 
\begin{eqnarray}\label{boscovder}
{\cal D}_\m\phi^A&=&\partial_\mu \phi^A
    - b_\mu\kild{A} + \ft12
{V_\mu}^{\!i}{}_k\,\varepsilon^{jk}k_{ij}^A - g\, W_\m^I\,k_I^A \,,
\nonumber\\  
{\cal D}_\m A_i{}^\a &=& \partial_\m A_i{}^\a -
b_\m A_i{}^\a +\ft12V_{\m i}{}^jA_j{}^\a  -g\,
W_\m{}^{\!\!\a}{}_{\!\!\b}\, A_i{}^\b \ +  
\pa_\m\phi^A\,{\G_A}^{\!\a}{}_{\!\b} \, A_i{}^\b\,, \\
{\cal D}_\m\z^\a&=&\partial _\m\z^\a -\ft14\o^{ab}_\m\,\g_{ab}
    \z^\a -\ft32b_\m\,\z^\a+\ft{1}{2}i A_\m\,\z^\a -g\,
W_\m{}^{\!\!\a}{}_{\!\!\b}\,\z^\b\,  
+\partial_\mu\phi^A\,{\G_A}^{\!\a}{}_{\!\b}\,\z^\b\ , \nonumber
\end{eqnarray}
where we have also included the terms related to possible gauged
isometries. All covariantizations follow straightforwardly from 
the formulae presented in section~3.3 and from the gauge field
conventions given in \cite{DVV,DWLVP}. Observe that the derivative in
$\pa_\m\phi^A$ multiplying the connection $\G_A{}^{\!\a}{}_{\!\b}$
does {\it not} require an additional covariantization. 

The transformation rules under $Q$- and $S$-supersymmetry are now as
follows, 
\bea
\d\phi^A&=& 2( \g^A_{i\bar\a} \,\bar\e^i 
\zeta^{\bar \a} +  
\bar\g^{Ai}_{\a} \,\bar\e_i \zeta^\a )\,,\nonumber\\
\d A_i{}^\a &=&  2\, \bar\e_i\zeta^\a + 2\, 
\varepsilon_{ij} \,
G^{\a\bar\b}\Omega_{\bar\b\bar\gamma} \,
\bar\e^j\zeta^{\bar\gamma}   - \d_{\scriptscriptstyle\rm Q}  
\phi^B \G_B{}^{\!\a}{}_{\!\b} \,A_i{}^\b \,,\nonumber \\
\d \zeta^\a &=& D\!\slash \,A_i{}^\a \e^i- 
\d_{\scriptscriptstyle\rm Q} \phi^B \G_B{}^{\!\a}{}_{\!\b}\, \zeta^\b 
 + 2g\, X^{\a}{}_{\!\b}\,A_i{}^\b \,
\varepsilon^{ij}\e_j  + A_i{}^\a\,\eta^i\,,\nonumber\\ 
\d \zeta^{\bar \a}  &=& D\!\slash\, 
A^{i\,\bar \a} \e_i- \d_{\scriptscriptstyle\rm Q}  
\phi^B \G_B{}^{\!\bar\a}{}_{\!\bar\b}\,\zeta^{\bar\b}  +2g\,\bar 
X^{\bar\a}{}_{\!\bar \b}\, A^{i\,\bar\b}\,
\varepsilon_{ij}\,\e^j +  A^{i\, \bar \a}\,\eta_i
\,,\label{local-QS-transf} 
\eea
where we have made use of the supercovariant derivatives (we also give
the supercovariant derivative of $\zeta^\a$ which is not needed
above),  
\bea 
D_\mu\phi^A &=& {\cal D}_\mu \phi^A
        - \g_{i{\bar \a}}^A\,{\bar \psi}_\mu{}^i\z^{\bar \a}
    - {\bar \g}_{\a}^{Ai}\,{\bar \psi}_{\mu\,i}\z^\a,\nonumber\\
D_\m A_i{}^\a&=&{\cal D}_\m A_i{}^\a-{\bar \psi}_{\m i}\z^\a-
\varepsilon_{ij}\,G^{\a\bar \b}\Omega_{\bar \b \bar \g}\,{\bar
\psi}_\m^j\z^{\bar \g}\nonumber\\
D_\mu\z^\a &=&
    {\cal D}_\mu\z^\a-\ft12 D \!\slash\, A_i{}^\a \psi_{\mu}^i
    -\ft12 A_i{}^\a \phi_{\mu}^i\,.
\eea
We have verified that no further modifications of the fermionic
transformation rules beyond those given above are possible, assuming
that the bosonic transformation rules remain the same. One of the
underlying reasons for the absence of additional terms may be that the
above rules are already consistent 
with rigid supersymmetry and with the case of a flat hyper-\Ka\
manifold which was
taken as a starting point in \cite{DWLVP}. All additional
modifications would thus have to vanish in the corresponding limits,
while at the same time one must preserve covariance under target-space  
diffeomorphisms and fermionic frame reparametrizations. Therefore the
possible modifications should be proportional to the target-space
curvature times the superconformal fields and, as it turns out,  it is
difficult if not impossible to see how such terms could emerge.
Given the fact that the transformation rules take the same form, we
expect the same situation for the Lagrangian, where, again, it is
difficult to construct suitable modifications that would vanish in the
appropriate limits. 

Motivated by these considerations, we write down the Lagrangian by
converting and covariantizing the relevant equation (3.28) in
\cite{DWLVP}. Here we suppress the hypermultiplet auxiliary fields, as
we no longer insist on off-shell supersymmetry for the
hypermultiplets. The result reads as follows, where the derivatives
are all fully covariantized, 
\bea
e^{-1} {\cal L} &=& \ft12 \varepsilon^{ij}\, \bar \Omega_{\a\b} \,
A_i{}^\a (D^aD_a + \ft32 D) A_j{}^\b \nonumber \\
&&+\bar\Omega_{\a\b} \Big[2\, g^2\, \varepsilon^{ij}\,A_i{}^\a \,\bar
X^\b{}_\g\,X^\g{}_\d\,A_j{}^\d + \ft12 g\,A_i{}^\a\,Y^{ij\b}{}_\g
A_j{}^\g \Big] \nonumber \\
&& - \Big[ ( \bar \zeta^\a - \ft12 \bar \psi^i _\m\g^\m
A_i{}^\a)\nonumber \\
&&\hspace{5mm}\times ( G_{\bar\b\a} 
D\!\slash\, \zeta^{\bar \b}  + \bar \Omega_{\a\b}(\ft32 \varepsilon^{jk}
\chi_j\,A_k{}^\b -\ft14 \varepsilon^{jk} 
\,T_{abjk} \,\g^{ab} \zeta^\b) \nonumber \\
&&\hspace{10mm}  - g\,\bar\Omega_{\a\b} (\Omega^{i^\b}{}_\g\,A_i{}^\g +2\,\bar
X^\b{}_\g \,\zeta^\g ) 
+ \mbox{ h.c.} \Big]  \nonumber\\
&&+\ft12 g \Big[ \bar\Omega_{\a\b} \,A_i{}^\a \,
\bar\Omega^{i\b}{}_\g \,\zeta^\g + \mbox{ h.c.}\Big] \,. 
\eea
After substituting the expressions for the dependent gauge fields
$\phi_\m^i$ and $f_\m^{\;a}$ in terms of the other
fields and dropping a total derivative, we write the Lagrangian as
follows, 
\begin{eqnarray}\label{action}
e^{-1}{\cal L}&=&-\ft12G_{\bar \a \b}{\cal D}_\m A_i{}^\b
{\cal D}^\m A^{i\bar \a}+\ft{1}{12}R\,G_{\bar \a \b}A_i{}^\b A^{i\bar \a}
+\ft14D\,G_{\bar \a \b}A_i{}^\b A^{i\bar \a }  \nonumber\\
&&-G_{{\bar \a}\b}\left({\bar \z}^{\bar \a}{\cal D} \!\slash\, \z^\b+{\bar
\z}^\b{\cal D} \!\slash\, \z^{\bar \a}\right) -\ft14  W_{\bar
\a\b\bar\g\d}\, \bar \zeta^{\bar \a}  
\g_\m\zeta^{\b}\,\bar \zeta^{\bar \g} \g^\m\zeta^\d 
\nonumber\\
&&+\Big[G_{\bar \a \b} \Bigl(-\ft{1}{12}A_i{}^\b A^{i\bar \a} 
 e^{-1}\varepsilon^{\m\n\rho \s}{\bar \psi}_{\m j}\g_\n
{\cal D}_\rho \psi_\s^j+\ft18 A_i{}^\b A^{i\bar \a}
 \,{\bar \psi}_{j\m}\g^\m \chi^j\nonumber\\
&&\quad-\ft{1}{48}A_k{}^\b A^{k\bar \a}\,
{\bar \psi}^i_\mu \psi^j_\nu\,T^{\mu\nu}_{ij}
-A^{i\bar \a}{\bar \z}^\b\chi_i+\ft1{16} {\bar \O}^{\bar \a \bar \g}
\,G_{\bar \g \l}\,{\bar \z}^\b \g^{ab}\,T_{abij}\varepsilon^{ij}\z^\l
\nonumber\\
&&\quad+{\bar \z}^\b \g^\m {\cal D} \!\slash\, A^{i\bar \a} \psi_{\m i}
-\ft23  A^{i\bar \a}\,{\bar \z}^\b\,\g^{\m \n}{\cal D}_\m \psi_{\n i}
+\ft{1}{24} A^{i\bar \a}\,{\bar \z}^\b \g^{ab} T_{abij}\g^\m
\psi^j_\m\nonumber\\
&&\quad-\ft14e^{-1}\varepsilon^{\m\n\rho \s}\,{\bar \psi}^i_\m \g_\nu
 \psi_{\rho j}\, A_i{}^\b 
{\cal D}_\s A^{j\bar \a}\nonumber\\
&&\quad-\ft12 {\bar \z}^\b \g^\m\g^\n\psi_{\m
i}({\bar \psi}^i_\n \z^{\bar \a}+\varepsilon^{ij}\,
{\bar \O}^{\bar \a \bar \rho}\,G_{\bar \rho \l}\,{\bar
\psi}_{\n j}\z^\l)+\mbox{h.c.}\Bigr) \Bigr]\ .
\end{eqnarray}
Here we did not include the terms related to gauged isometries. To
incorporate those one includes the relevant terms into the covariant
derivatives and adds the following $g$-dependent terms to the
Lagrangian, 
\begin{eqnarray}
e^{-1}{\cal L}_g&=&2g^2G_{\bar \a \b}\, A^{i\bar \a}\,\bar
X^\b{}_\g\,X^\g{}_\delta \,A_i^\delta  +
\ft12 g\,A_i{}^\a\, {\bar \O}_{\a\b}\,Y^{ij\,\b}{}_{\!\g} 
\, A_j{}^\g   \nonumber\\
&& +g\,\Big[\,2\, {\bar X}^\g{}_\a\,{\bar
\z}^\a\z^\b\, {\bar \O}_{\b \g}  +2\,{\bar \O}_{\a\b}\,
{\bar \z}^\a\,\O^{i\b}{}_\delta\,A_i{}^\delta
\nonumber\\ 
&&\hspace{7mm} -2\,{\bar \psi}^i_\m\g^\m \z^\b \,{\bar X}^\a{}_\b\,
{\bar \O}_{\a \g}\,A_i{}^\g
-\ft12 \,{\bar \psi}^i_\m \g^\m \Omega^{k\a}{}_\b\, {\bar \O}_{\a
\g}\,A_i{}^\g \,A_k{}^\b\nonumber\\
&&\hspace{7mm} -\ft12{\bar \psi}^i_\m\g^{\m\n}\psi_\n^k\, A_k{}^\b\,
{\bar\O}_{\a \g} \,A_i{}^\g
{\bar X}^\a{}_\b+\mbox{h.c.}\,\Big]\ .\label{gauged-action}
\end{eqnarray}
As mentioned above, these results are in agreement with
the action presented in subsection~2.2 as well as with the results of
\cite{DWLVP} in the appropriate limits. In addition we performed a
number of 
independent checks on \eqn{action} and \eqn{gauged-action}. 
For instance, because the superalgebra closes only modulo the field
equations for the fermion fields $\zeta^\a$ and $\zeta^{\bar\a}$, we
have calculated these field equations from the supersymmetry
transformation rules \eqn{local-QS-transf}. As it turns out the result is in
agreement with the field equations derived from the action. 

The above action is invariant under all superconformal symmetries. In
particular the scalar fields are subject to dilatations and to SU(2)
transformations. Ignoring the contributions from the vector
multiplets, which are essential for obtaining the complete and
consistent action for Poincar\'e supergravity coupled to vector
multiplets and hypermultiplets, but which do not affect the
target-space geometry of the hypermultiplets, we express the bosonic
terms in scale-invariant quantities, by introducing a normalized
section 
\be
\hat A_i{}^\a = \chi^{-1/2} \, A_i{}^\a  \,,
\ee
which satisfies $\Omega_{\a\b} \,\hat A_i{}^\a\,\hat A_j{}^\b =
\varepsilon_{ij}$. Similarly we redefine the various other fields,
such as the vierbeine, spin connection, etcetera, by
a $\chi$-dependent scale transformation. The result for the bosonic
terms then takes the form
\be
{\cal L} = -\ft12 e\Big[\bar\Omega_{\a\b} \,\varepsilon^{ij}  \,{\cal
D}_\m\hat A_i{}^\a\,{\cal D}^\m \hat A_j{}^\b - \ft 13 R -   D\Big] \,,
\ee
where $R$ is the Ricci scalar of the spacetime.
Suppressing  possible gauged isometries for convenience, this results in
\bea
{\cal L} &=& -\ft12 e \,\bar\Omega_{\a\b} \,\varepsilon^{ij}
\,  (\pa_\m\phi^A\,D_A\hat A_i{}^\a +\ft12 V_{\m i}{}^k
\,\hat A_k{}^\a) \,(\pa^\m\phi^B\, D_B \hat A_j{}^\b +\ft12
V^\m_{j}{}^l \hat A_l{}^\b) \nonumber \\ 
&& + \ft 16 e\, R + \ft12 e\, D \,.
\eea
The field equations for the SU(2) gauge fields $V_{\m i}{}^j$ 
yield, 
\be 
V_{\m i}{}^j= -2\,\pa_\m \phi^A \,{\cal V}_{A\,ik} \,\varepsilon^{kj} \,.
\ee
This result can be substituted back into the Lagrangian, which then
reads 
\be
{\cal L} = -\ft12 e \, G_{AB} \,\pa_\m\phi^A\,\pa^\m\phi^B
 + \ft 16 e\, R + \ft12 e\, D \,,
\ee
so that the target-space metric $G_{AB}$ corresponds indeed to the
quaternionic space which we constructed in the previous section. The
terms with 
the Ricci scalar and the auxiliary field $D$ combine with similar terms
from the Lagrangian of the vector multiplets to give the
Einstein-Hilbert action. 

The material derived in the previous section now fits in nicely
with what is known about the general coupling of hypermultiplets to
supergravity \cite{BagWit}. First of all, the quantity  
$\Delta_{\a\b}$ projects out precisely the $S$-invariant hypermultiplet
spinors which thus describe $2n-2$ physical spinors after modding
out the $S$-supersymmetry. Hence, the nonlinear sigma model comprises
precisely the expected $4n-4$ scalars and $2n-2$ spinors. The relevant
quaternionic vielbeine have already been defined in
\eqn{def-reduced-vielbeine},  but can 
equally well be obtained from working out the above Lagrangian after
removing the appropriate gauge degrees of freedom 
We will list a number of relevant
identities, which all follow from the previous section, 
\bea
\Delta_{\a\b}\, \hat V_{Ai}^\a\,\hat V_{Bj}^\b &=& \ft12
\varepsilon_{ij} \,G_{AB} + \ft12 R_{ABij} \,,\nonumber \\
G_{AB} \,\bar{\hat \g}^{Ai}_\a\,\bar{\hat\g}^{Bj}_\b  &=&
\varepsilon^{ij}\,\Delta_{\a\b} \,,\nonumber  \\
R_{ABij}\,\bar{\hat\g}^{Ak}_\a\,\bar{\hat\g}^{Bl}_\b &=&
2\, \d_i^{(k}\d_j^{l)}\,\Delta_{\a\b}\,.
\eea
The second integrability condition \eqn{integrab2} can be rewritten as
\be
\hat R_{ABCD} \,\bar{\hat\gamma}{}^{Ci}_\a
\,\bar{\hat\gamma}{}^{Dj}_\b = -\varepsilon^{ij}\,\Delta_{\a\g}\,\hat
R_{AB}{}^{\!\g}{}_{\!\b} - \Delta_{\a\b} \,R_{AB}^{ij}\,,
\ee
which gives the decomposition of the Riemann tensor into an Sp$(n-1)$
and an Sp(1) curvature. Of course, this relation is already
incorporated into the expression \eqn{curvs-into-W} and its
correctness can also be verified directly. 
The curvature $\hat R_{AB}{}^{\!\a}{}_{\!\b}$ satisfies
(c.f. \eqn{new-sympl-curv}),
\be
\bar\Omega_{\a\g}\,{\hat R}_{AB}{}^{\!\g}{}_{\!\b} =
\bar\Omega_{\a\g}\,R_{AB}{}^{\!\g}{}_{\!\b} - 2 \varepsilon_{ij}
\,\bar{\hat\gamma}{}^i_{A(\a} \,\bar{\hat\gamma}{}^j_{B\b)} +
\bar\Omega_{\a\g} \,\bar\Omega_{\b\d} \,\hat A_i{}^\g\,A_j{}^\d
\,R_{AB}^{ij}\,.
\ee
Upon projection with $\Delta$, the last term vanishes and one finds an
identity that is well-known from the literature. 
 
Hence we see that all aspects of quaternionic geometry that arise in
the coupling of hypermultiplets to supergravity are correctly
reproduced. Our results provide an elegant extension of the work
reported in \cite{DWLVP} and give a unified prescription for all
hypermultiplet couplings to supergravity. Although this is in principle
straightforward, it remains to work out the details of the Lagrangian
and transformation rules after removing the gauge degrees of freedom
associated with $S$-supersymmetry. 

\vspace{8mm}

\noindent
{\bf Acknowledgement}\\
We acknowledge useful discussions with J. Figueroa-O'Farrill,  
G. Gibbons and A. Swann. B.d.W. is grateful to
the Alexander von Humboldt-Stiftung  for supporting his stay at the
AEI as part of the Humboldt Award program.
S.~V. thanks PPARC for financial support during his stay at the University
of Swansea, and the Institute for 
Theoretical Physics in Utrecht and the AEI for their hospitality.
This work is supported in part  by the European Commission TMR
programmes FMRX-CT96-0012, in which the Albert Einstein Institut and
the University of Wales in Swansea participate, and 
ERBFMRX-CT96-0045, in which Utrecht University participates. 
%

\end{document}